\documentclass[12pt, peerreview , onecolumn]{IEEEtran}
\usepackage{tikz}
\usepackage[sumlimits,intlimits]{amsmath}
\usepackage{amsfonts,amssymb}%
\usepackage{bm}%
\usepackage{graphicx,graphics}%
\usepackage[noadjust]{cite}%
\usepackage{color}%
\usepackage{url}
\usepackage{verbatim}
\usepackage{balance}
\usepackage{multirow}
\usepackage{multicol}
\usepackage{stfloats}
\usepackage{xspace}
\usepackage{enumerate}
\usepackage{rotating}
\usepackage[tight]{subfigure}
\usepackage{algorithm}
\usepackage{algorithmicx, algpseudocode}

\newtheorem{proposition}{Proposition}

\newcommand{\redcom}[1]{{\color{red}#1}\xspace}

\newcommand{\bh}{\bar{h}}
\newcommand{\bg}{\bar{g}}

\newcommand{\bP}{\bar{P}}
\newcommand{\bp}{\bar{p}}
\newcommand{\bt}{\bar{t}}
\newcommand{\ba}{\bar{\alpha}}
\newcommand{\bv}{\bar{v}}
\newcommand{\bu}{\bar{u}}

\newtheorem{remark}{Remark}

\ifCLASSOPTIONpeerreview
\renewcommand{\baselinestretch}{2.0}
\fi

\begin{document}

%\ifCLASSOPTIONpeerreview
%    \title{\vspace*{-0pt}{Coordinated Scheduling and Power Control for }
%\\ \vspace*{-20pt}
%heterogeneous multicell networks with
%\\ \vspace*{-20pt}
%Wireless Energy Harvesting}
%\else
%%\title{Wireless Energy Harvesting and Information Relaying: Adaptive Time-Switching Protocols and Throughput Analysis}
%
%\title{\vspace*{-0pt}{Coordinated Scheduling and Power Control for heterogeneous multicell networks with Wireless Energy Harvesting}}
%
%\fi

\title { \vspace*{-0pt} {\redcom{Joint Resource Optimization for Multicell Networks}}
\\ \vspace*{-20pt}
with Wireless Energy Harvesting Relays}
%\\ \vspace*{-20pt}

%\title{Joint Resource Optimization for Heterogeneous Multicell Networks with Wireless Energy Harvesting Relays}

\author{Ali A.~Nasir, Duy~T.~Ngo, Xiangyun~Zhou, Rodney~A.~Kennedy, and Salman~Durrani%
\thanks{Ali A. Nasir, Xiangyun Zhou, Rodney A. Kennedy and Salman Durrani are with the Research School of Engineering, the Australian National University, Canberra, ACT 2601, Australia (Email: \{ali.nasir, xiangyun.zhou, rodney.kennedy, salman.durrani\}@anu.edu.au). Duy T. Ngo is with the School of Electrical Engineering and Computer Science, the University of Newcastle, Callaghan, NSW 2308, Australia (Email: duy.ngo@newcastle.edu.au).}
}

\maketitle
\thispagestyle{empty}
\pagestyle{empty}

\vspace{-12pt}

\begin{abstract}
This paper \redcom{first} considers a multicell network deployment where the base station (BS) of each cell communicates with its cell-edge user with the assistance of an amplify-and-forward (AF) relay node. Equipped with a power splitter and a wireless energy harvester, the self-sustaining relay scavenges radio frequency (RF) energy from the received signals to process and forward the information. Our aim is to develop a resource allocation scheme that jointly optimizes (i) BS transmit powers, (ii) received power splitting factors for energy harvesting and information processing at the relays, and (iii) relay transmit powers. In the face of strong intercell interference and limited radio resources, we formulate three highly-nonconvex problems with the objectives of sum-rate maximization, max-min throughput fairness and sum-power minimization. To solve such challenging problems, we propose to apply the successive convex approximation (SCA) approach and devise iterative algorithms based on geometric programming and difference-of-convex-functions programming. The proposed algorithms transform the nonconvex problems into a sequence of convex problems, each of which is solved very efficiently by the interior-point method. We prove that our algorithms converge to the \redcom{locally} optimal solutions that satisfy the Karush-Kuhn-Tucker conditions of the original nonconvex problems. \redcom{We then extend our results to the case of decode-and-forward (DF) relaying with variable timeslot durations. We show that our resource allocation solutions in this case offer better throughput than that of the AF counterpart with equal timeslot durations, albeit at a higher computational complexity.} Numerical results confirm that \redcom{the proposed} joint optimization solutions substantially improve the network performance, compared with cases where the radio resource parameters are individually optimized.
\end{abstract}

%  and schedule simultaneous transmission
\ifCLASSOPTIONpeerreview
    \newpage
\fi

%\vspace{-7pt}
%
%\ifCLASSOPTIONpeerreview
%    \newpage
%\fi 

\section{Introduction}

\redcom{Multicell networks with universal frequency reuse play an important role in meeting the ever increasing demand of ubiquitous wireless coverage and high data throughput in the near future \cite{Andrews-14-Jun-A,Hwang-Jun-13-A,Sambo-14-A}. One of the challenges in such networks is to maintain the quality of service requirements for cell-edge users due to the interference from the neighboring cells \cite{Hwang-Jun-13-A,Andrews-14-Jun-A}. The deployment of relays is regarded as a viable solution in eliminating coverage holes in areas that are otherwise difficult for BSs' signals to penetrate \cite{Perez-11-A,Yang-09-Oct-A}. In addition, the performance of multicell networks can be further enhanced by utilizing coordinated multipoint transmission and reception (CoMP) techniques \cite{Irmer-11-Feb-A,Yang-13-A}, in which BSs and relays cooperate with one another to best serve the cell-edge users.

Due to random positions and mobility of users, relays need to be opportunistically deployed where most needed. This can be achieved if relays do not require a wired power connection and are powered using alternative `green' energy resources. Recently, radio frequency (RF) or wireless energy harvesting has emerged as an attractive solution to power wireless nodes~\cite{Lu-14-A}. While energy harvesting from ambient sources may not be sufficient to power relay nodes, carefully designed wireless power transfer links can be used to power relay nodes~\cite{Lu-14-A,Tabassum-15,Huang-15-A}. In this regard, it is crucial to ensure that the very different information decoding and power transfer power sensitivity requirements are met at the receiver (e.g., $-60$ dBm for information receivers and $-10$ dBm to $-30$ dBm for energy receivers~\cite{Lu-14-A}).

A multicell network with energy harvesting relays poses interesting design challenges, such as: (i) How to effectively manage intercell interference, (ii) How to allocated limited power at the base stations (BSs), (iii) How to design wireless power transfer links for amplify-and-forward (AF) and decode-and-forward (DF) relays, and (iv) How the harvested RF energy is utilized at the relays. Existing research in the literature has partially addressed these important issues. The design of wireless energy harvesting relays in point-to-point single-cell systems is considered in \cite{Ding-13-A,Ding-A2-14,Ding-13-Dec-A,Cumanan-13-Oct-A,Nasir-13-A,Nasir-14-P,Nasir-15-A}. Assuming simultaneous wireless information and power transfer in a single-cell network, the power control problem for multiuser broadband wireless systems without relays is studied in \cite{Huang-13-A}. In \cite{Wang-14-P}, a similar problem is examined, albeit in the context of multiuser multi-input-multi-output (MIMO) systems. Considering relays in a single-cell network, resource allocation schemes for the remote radio heads are specifically developed in \cite{Ng-14-P}. In the downlink of a multicell multiuser interference network, coordinated scheduling and power control algorithms for the macrocell BSs only are proposed in \cite{Venturnio-09-A,Ksairi-11-A}. Recently, in \cite{Chen-P-14}, an optimal power splitting rule is devised for energy harvesting and information processing at the self-sustaining relays of multiuser interference networks. However, \cite{Chen-P-14} does not consider the important issue of allocating the transmit powers at the BSs and the relays.

In this paper, we consider a multicell network in which the BS of each cell communicates with its cell-edge user via a wireless energy harvesting relay node. The relay is equipped with an energy harvesting receiver and information transceiver. We assume that the energy harvesting receiver implements a power-splitting (PS) based receiver architecture \cite{Zhou-12-A}, i.e., the relay uses a portion of the received signal power for energy harvesting and the remaining signal energy as input to the information transceiver. Using the harvested energy, the information transceiver employs either AF or DF relaying to forward the received signal to its corresponding user. The BSs in the multicell network adopt CoMP, i.e., they share the channel quality measurements and schedule the transmissions, allowing for more efficient radio resource utilization.

First, we formulate three new resource optimization problems for multicell networks with EH-enabled AF relays, namely, sum-rate maximization, minimum-throughput maximization, and sum-power minimization\footnote{ \redcom{ A preliminary version of this work, which considers the sum-rate maximization problem for AF relaying only, has been accepted for presentation at the 2015 IEEE International Conference on Communications (ICC), London, U.K. \cite{Nasir-15-ICC-P}.}}. The objective is to jointly optimize the transmit powers at the BSs and the relays and also find the optimal power splitting rule at the relays. Our formulations directly target the critical issue of multicell interference, at the same time as meeting the stringent constraints on the available transmit powers at the BSs and the relays. Since the optimization variables are strongly coupled with many nonlinear cross-multiplying terms, the formulated problems are highly nonconvex. To the best of our knowledge, there exists no practical method that guarantees to offer the true global optimality to these challenging problems.

Then, we exploit the problem structure and adopt the successive convex approximation (SCA) method to transform the highly nonconvex problems into a series of convex subproblems. Here, we specifically tailor the generic SCA framework via the applications of geometric programming (GP) and difference-of-convex-functions (DC) programming. At each step of our proposed iterative algorithms, we efficiently solve the resulting convex problem by the interior-point method. We analytically prove that our developed algorithms generate a sequence of improved feasible solutions, which eventually converge to a locally optimal solution satisfying the Karush-Kuhn-Tucker (KKT) conditions of the original problems. Note that the general convergence analysis of SCA method is established in \cite{Razaviyayn-13-A} and SCA-based solutions have been empirically shown to often achieve the global optimality in many practical applications, e.g., in wireline DSL networks \cite{PapanEvans-09}, wireless interference networks \cite{Chiang-07-A,Kha-12-A}, and small-cell heterogeneous networks \cite{Ngo-14-A}.

Finally, we show that the proposed SCA-based approach can be extended to the more general case of variable timeslot durations with DF relaying. Numerical examples with realistic network parameters confirm that our joint optimization solutions significantly outperform those where the radio resource parameters are individually optimized.
%\end{itemize}

The rest of this paper is organized as follows: Sec. \ref{sec:sys_mod} presents the system model and states the key assumptions used throughout this work. Sec. \ref{sec:SM_AF} presents the signal model for AF relaying and equal timeslot durations. Sec. \ref{sec:PF} formulates the nonconvex resource allocation problems and introduces the generic SCA framework. Secs. \ref{sec:GP} and \ref{sec:DC} propose the GP-based and DC-based SCA solutions for AF relaying, respectively. Sec. \ref{sec:DF} extends our results to the case of variable timeslot durations with DF relaying. Sec. \ref{sec:sim} presents numerical results to confirm the advantages of our proposed algorithms. And Sec. \ref{sec:conclusions} concludes the paper.}

\section{System Model and Assumptions}\label{sec:sys_mod}

\begin{figure}[t]
    \centering
    \includegraphics[width=0.8 \textwidth]{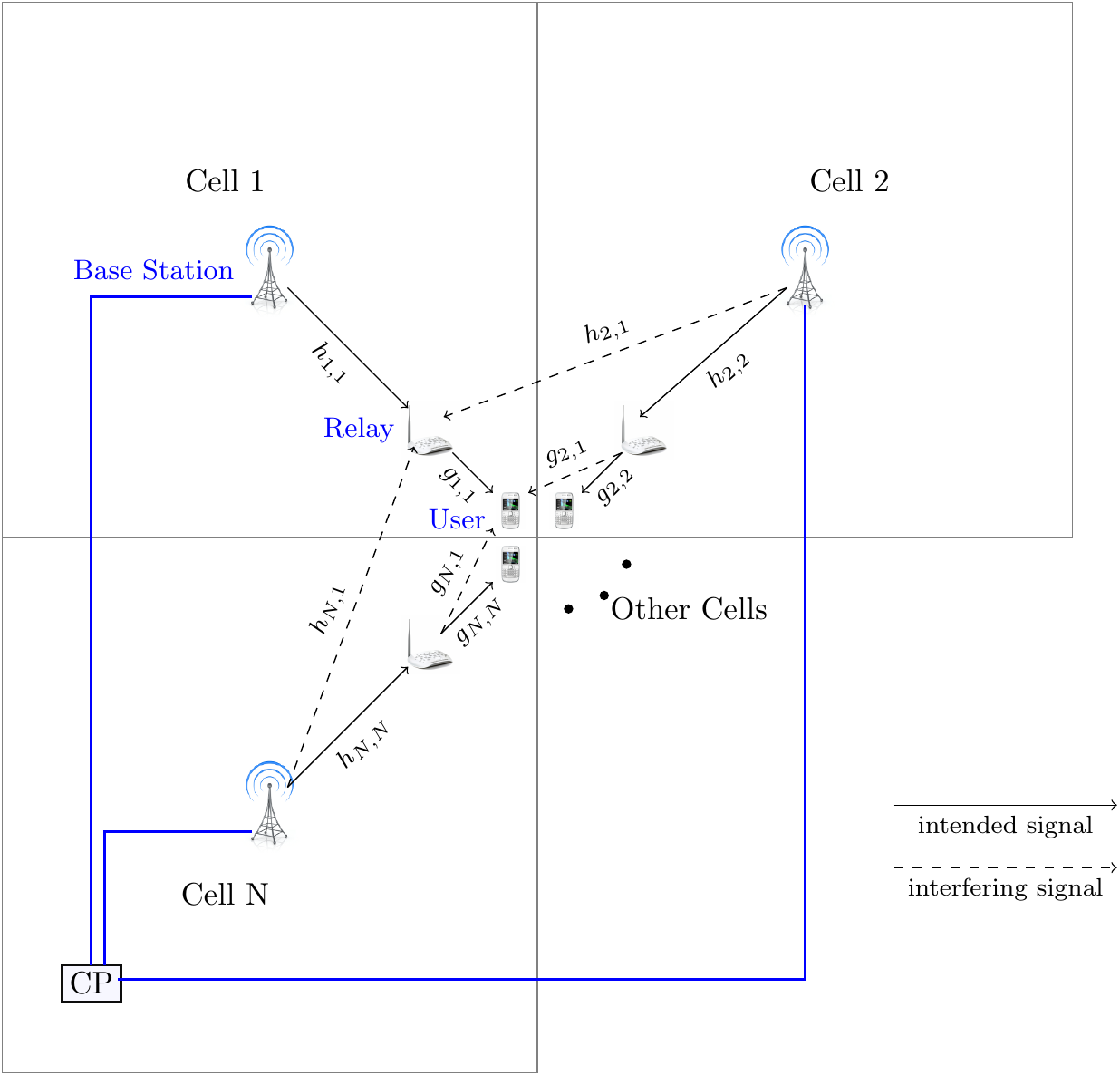} \vspace{-.15 in}
  \caption{A multicell network consists of $N$ cells and a central processing (CP) unit. Each cell has a base station, a relay and a cell-edge user. For clarity, we only show the interfering scenarios in Cell 1, i.e., at relay $1$ and user $1$. In general, interference happens at all $N$ relays and $N$ users.}
    \label{fig:sys_mod}
\end{figure}

\redcom{Consider the downlink transmissions in an $N$-cell network with universal frequency reuse, i.e., the same radio frequencies are used in all cells. Adopting CoMP, we assume that the base stations (BSs) are connected to a central processing (CP) unit which coordinates the multicellular transmissions and radio resource management. The network under consideration is illustrated in Fig. \ref{fig:sys_mod}. Note that although square-cells are shown in Fig. \ref{fig:sys_mod}, the analysis and proposed solutions in this paper are valid for any cellular network geometry.

Let $\mathcal{N} = \{1,\hdots,N\}$ denote the set of all cells. In each cell $i\in\mathcal{N}$, the BS attempts to establish communication with its cell-edge users. We assume that these users are located in the `signal dead zones', where no direct signal from their serving BS can reach. A relay node is deployed in each cell to assist in forwarding the signal from the BS, extending the network coverage to the distant users. We assume that orthogonal channels are assigned to users in each cell (e.g., by means of TDMA, FDMA or OFDMA); hence, the intracell interference is eliminated. Therefore, we only focus on the resource allocation in one channel, which corresponds to only one user in a cell. By BS $i$, relay $i$ and user $i$, we mean the BS, the relay and the single user of cell $i\in\mathcal{N}$, respectively.

We assume that the relays are energy-constrained nodes and they harvests energy from the RF signals of all BSs, using the power-splitting based receiver architecture. While each BS has a maximum power limit $P_\text{max}$ available for transmission, it must transmit with a minimum transmit power $P_\text{min}$ to ensure that the energy harvesting circuit at the relay is activated. The harvested energy is used by a relay transceiver to process and forward the BS signal to its intended user. We further assume that the relays are mounted on the building rooftops to have a line-of-sight link from the serving BSs.

Let $h_{i,j}$ be the channel coefficient from the BS $i$ to relay $j$ and $g_{j,k}$ be the channel coefficient from the relay $j$ to user $k$. We assume that all the BSs send the available channel state information (CSI) to the CP unit via a dedicated control channel. In this paper, we assume perfect knowledge of CSI at the BSs, allowing for a benchmark performance to be determined.}

\section{Signal Model with AF relaying}\label{sec:SM_AF}

\begin{figure}[t]
    \centering
    \includegraphics[width=1.0 \textwidth]{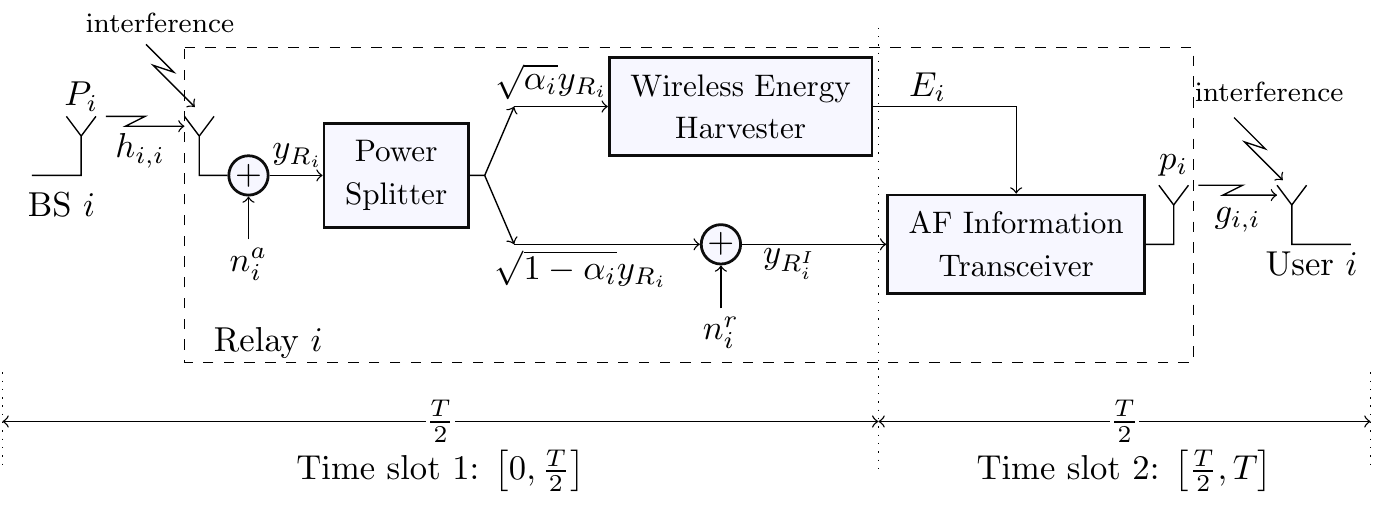} \vspace{-.15 in}
  \caption{BS-to-user communication assisted by a wireless energy harvesting AF relay.}
    \label{fig:relay}
\end{figure}

\redcom{We first consider the case of AF relaying where we divide the total transmission block time $T$ into two equal timeslots. The first timeslot includes BS-to-relay transmissions and energy harvesting at the relays. During the first timeslot, the relays do not transmit. The second timeslot includes signal processing at the relays and relay-to-user transmissions. In this second timeslot, the BSs do not transmit. The operations in each timeslot are illustrated in Fig.~\ref{fig:relay}, which will be further discussed in the following.}

%In the first timeslot, the BSs transmit to the relays while all the relays do not transmit. At the relay receivers, RF energy is scavenged by wireless energy harvesters. In the second timeslot, using the harvested energy, relay transmitters amplify and forward the information signals by the BSs to the respective users. This is done while all the BSs suppress their transmissions. In what follows, we will describe the detailed operations in each timeslot.

%As shown in Fig. \ref{fig:relay_bd},

\subsection{BS-to-Relay Transmissions and Wireless Energy Harvesting at Relay Receivers}

In the first timeslot $[0,T/2]$, let $x_i$ be the normalized information signal to be sent by BS $i$, i.e., $\mathbb{E} \{ |x_i |^2 \} = 1$, where $\mathbb{E}\{\cdot\}$ denotes the expectation operator and $|\cdot|$ the absolute value operator. Let $P_\text{min} \le P_i \le P_\text{max}$ denote the transmit power of BS $i$, $d_{i,j}^h$ the distance between BS $i$ and relay $j$, and $\beta$ the path-loss exponent. Assuming that $n_i^a$ is the zero-mean additive white Gaussian noise (AWGN) with variance $\sigma_i^a$ at the receiving antenna of relay $i$, the received signal at relay $i$ can be expressed as:
\begin{align}\label{eq:yr}
      y_{R_i} =  \frac{h_{i,i}}{\sqrt{ \left( d_{i,i}^h \right)^\beta}} \sqrt{P_i}  x_i  +  \sum_{j=1,j \ne i}^{N} \frac{h_{j,i}}{\sqrt{ \left(d_{j,i}^h \right)^\beta}} \sqrt{P_j}  x_j + n_i^a.
\end{align}
%

\begin{comment}
\begin{table}[t]
%\centering
\caption{List of commonly used symbols.} \centering
\begin{tabular}{|l|l|l|l|} \hline
\multicolumn{1}{|c|} {Symbol} & \multicolumn{1}{c|} {Meaning} \\ \hline
$P_i$ & Transmit power of BS $i$. \\
$p_i$ & Transmit power of relay transceiver $i$. \\
$\alpha_i$ & Power splitting factor of relay $i$.  \\ \hline
\end{tabular}
\label{tab:1}
\end{table}
\end{comment}

\redcom{We assume that each relay is equipped with a power splitter that determines how much received signal energy should be dedicated to the energy harvester and the signal processing receiver \cite{Zhou-12-A,Chen-P-14,Ding-13-A,Ding-A2-14}}. As shown in Fig. \ref{fig:relay}, the power splitter at relay $i\in\mathcal{N}$ divides the power of $y_{R_i}$ into two parts in the proportion of $\alpha_i : (1-\alpha_i)$. Here, $\alpha_i \in (0,1)$ is termed as the power splitting factor. The first part $\sqrt{\alpha_i} y_{R_i}$ is processed by the energy harvester and stored as energy (e.g., by charging a battery at relay $i$) for the use in the second timeslot. The amount of energy harvested at relay $i$ is given by:
\begin{align}\label{eq:E}
      E_i =  \frac{\eta \alpha_i T}{2} \sum_{j=1}^{N} P_j \bh_{j,i},
\end{align}
where $\eta \in (0,1)$ is the efficiency of energy conversion and $\bh_{j,i} \triangleq  {|h_{j,i}|^2}{\left( d_{j,i}^h \right)^{-\beta}}, \ \forall i,j \in \mathcal{N}$, is the effective channel gain from BS $j$ to relay $i$ (including the effects of both small-scale fading and large-scale path loss).

The second part $\sqrt{1-\alpha_i} y_{R_i}$ of the received signal is passed to an information transceiver. %Note that due to the effect of intercell interference, the choice of $\alpha_i$ in a cell $i\in\mathcal{N}$ will eventually affect the throughput of other cells, and hence the overall network throughput performance.
In Fig. \ref{fig:relay}, $n_i^r$ denotes the AWGN with zero mean and variance ${\sigma_i^r}$ introduced by the baseband processing circuitry. Since antenna noise power ${\sigma_i^a}$ is very small compared to the circuit noise power ${\sigma_i^r}$ in practice \cite{Liu-13-A-DPST}, $n_i^a$ has a negligible impact on both the energy harvester and the information transceiver of relay $i$. Thus, for simplicity, we will ignore the effect of $n_i^a$ in the following analysis by setting $\sigma_i^a = 0$. The signal at the input of the information transceiver of relay $i$ can be written as:
\begin{align}\label{eq:yri}
      y_{R_i}^{I} &=  \sqrt{1 - \alpha_i} y_{R_i} + n_i^r =  \sqrt{1 - \alpha_i} \frac{h_{i,i}}{\sqrt{ \left( d_{i,i}^h \right)^\beta}} \sqrt{P_i}  x_i  +  \sqrt{1 - \alpha_i} \sum_{j=1,j \ne i}^{N} \frac{h_{j,i}}{\sqrt{ \left(d_{j,i}^h \right)^\beta}} \sqrt{P_j}  x_j + n_i^r,
\end{align}
where the first term in \eqref{eq:yri} is the desired signal from BS $i$, and the second term is the total interference from all other BSs.

\subsection{Signal Processing at Relays and Relay-to-User Transmissions}

In the second timeslot $[T/2, T]$, the information transceiver amplifies the signal $y_{R_i}^{I}$ prior to forwarding it to user $i$. Denote the transmit power of relay transceiver $i$ as $p_i$.  With the harvested energy $E_i$ in \eqref{eq:E}, the maximum power available for transmission at relay $i$ is given by $\frac{E_i}{T/2}=\frac{2E_i}{T}$, which means that:
\begin{align}\label{eq:pi}
      p_i \le \frac{2E_i}{T} = \eta \alpha_i \sum_{j=1}^{N} P_j \bh_{j,i}.
\end{align}
The transmitted signal from relay $i$ to user $i$ can then be written as:
\begin{align}\label{eq:xri}
      x_{R_i} = \frac{\sqrt{p_i} y_{R_i}^{I}}{\displaystyle\sqrt{(1-\alpha_i) \sum_{j=1}^{N} P_j \bh_{j,i} + \sigma_i^r  } },
\end{align}
where the denominator of \eqref{eq:xri} represents an amplifying factor that ensures power constraint \eqref{eq:pi} be met.
% Note that due to intercell interference, it is not always recommended to use the maximum power available for transmission at the relays. The choice of relay power, $p_i$, in the $i$th cell will eventually affect the throughput performance of other cells too.

Now, the received signal at user $i$ is:
\begin{align}\label{eq:yui}
      y_{U_i} =  \frac{g_{i,i}}{\sqrt{ \left( d_{i,i}^g \right)^\beta}}  x_{R_i}  +  \sum_{j=1,j \ne i}^{N} \frac{g_{j,i}}{\sqrt{ \left(d_{j,i}^g \right)^\beta}}   x_{R_j} + n_i^u,
\end{align}
where $d_{i,j}^g$ denotes the distance between relay $i$ and user $j$, and $n_i^u$ the AWGN with zero mean and variance $\sigma_i^u$ at the receiver of user $i$. Substituting $x_{R_i}$ in \eqref{eq:xri} into \eqref{eq:yui} yields:
\begin{align}\label{eq:yui_2}
      y_{U_i} =  \frac{g_{i,i} \sqrt{p_i} y_{R_i}^{I}}{\displaystyle\sqrt{\left( d_{i,i}^g \right)^\beta \left[(1-\alpha_i) \sum_{k=1}^{N} P_k \bh_{k,i} + \sigma_i^r\right]}}  +  \sum_{j=1,j \ne i}^{N} \frac{g_{j,i} \sqrt{p_j} y_{R_j}^{I}}{\displaystyle\sqrt{\left( d_{i,i}^g \right)^\beta \left[(1-\alpha_j) \sum_{k=1}^{N} P_k \bh_{k,j} + \sigma_j^r\right]}} + n_i^u.
\end{align}
With $y_{R_i}^{I}$ defined in \eqref{eq:yri}, we can then write \eqref{eq:yui_2} explicitly as:
\begin{align}\label{eq:yui_3}
      y_{U_i} &=  \frac{g_{i,i} h_{i,i} \sqrt{p_i P_i (1-\alpha_i) } x_i }{\displaystyle\sqrt{\left( d_{i,i}^g d_{i,i}^h \right)^\beta \left[(1-\alpha_i) \sum_{k=1}^{N} P_k \bh_{k,i} + \sigma_i^r \right]}} + \frac{\displaystyle g_{i,i} \sqrt{p_i (1-\alpha_i) } \sum_{j=1,j \ne i}^N  \frac{h_{j,i}}{\sqrt{ \left( d_{j,i}^h \right)^\beta}} \sqrt{P_j} x_j }{\displaystyle\sqrt{\left( d_{i,i}^g  \right)^\beta \left[(1-\alpha_i) \sum_{k=1}^{N} P_k \bh_{k,i} + \sigma_i^r \right]}} \nonumber\\
        & \quad +  \frac{g_{i,i}  \sqrt{p_i } n_i^r }{\displaystyle\sqrt{\left( d_{i,i}^g  \right)^\beta \left[(1-\alpha_i) \sum_{k=1}^{N} P_k \bh_{k,i} + \sigma_i^r \right]}}  + \sum_{j=1,j \ne i}^{N} \frac{g_{j,i} \sqrt{p_j} y_{R_j}^{I}}{\displaystyle\sqrt{\left(d_{j,i}^g \right)^\beta \left[(1-\alpha_j) \sum_{k=1}^{N} P_k \bh_{k,j} + \sigma_j^r \right]}} + n_i^u.
\end{align}
The first term in \eqref{eq:yui_3} represents the desired signal from BS $i$ to its serviced user $i$, whereas other terms represent the intercell interference and the noise.

Without loss of generality, let us assume $\sigma_i^r = \sigma_i^u = \sigma, \ \forall i \in \mathcal{N}$. The signal-to-interference-plus-noise ratio (SINR) at the receiver of user $i$ can be derived from \eqref{eq:yui_3} as:
\begin{align}\label{eq:sinr}
   \gamma_i = \frac{\phi_1^{i,i} P_i p_i (1-\alpha_i)}{\displaystyle\sum_{j=1,j \ne i}^N \phi_1^{i,j}  P_j p_i (1-\alpha_i) + \sum_{j=1}^N \left( \phi_2^{i,j}  P_j (1-\alpha_i) + \phi_3^{i,j}  p_j \right) +  \sum_{j=1,j \ne i}^N \sum_{k=1}^N \phi_4^{i,j,k} P_k p_j (1-\alpha_i) +1},
\end{align}
where we define
\begin{align}
      \phi_1^{i,j} \triangleq \frac{\bg_{i,i} \bh_{j,i}}{\sigma^2}; \quad \phi_2^{i,j} \triangleq \frac{\bh_{j,i}}{\sigma}; \quad \phi_3^{i,j} \triangleq \frac{\bg_{j,i} }{\sigma}; \quad \phi_4^{i,j,k} \triangleq \frac{\bg_{j,i} \bh_{k,i}}{\sigma^2}.
\end{align}
where $\bg_{j,i} \triangleq  {|g_{j,i}|^2}{\left( d_{j,i}^g \right)^{-\beta}}, \ \forall i,j \in \mathcal{N}$. For notational convenience, let us also define $\mathbf{P} \triangleq [P_1,\hdots,P_N]^T, \mathbf{p} \triangleq [p_1,\hdots,p_N]^T$, and $\boldsymbol\alpha \triangleq [\alpha_1,\hdots,\alpha_N]^T$. From \eqref{eq:sinr}, the achieved throughput in bps/Hz (bits per second per Hz) of cell $i$ is given by
\begin{align}\label{eq:tau}
   \tau_i(\mathbf{P},\mathbf{p},\boldsymbol\alpha) = \frac{1}{2} \log_2 (1 + \gamma_i).
\end{align}
An important observation from \eqref{eq:sinr} and \eqref{eq:tau} is that by dedicating more received power at relay $i$ for energy harvesting (i.e. increasing $\alpha_i$), one might actually decrease the end-to-end throughput in cell $i$. This can be verified upon dividing both the numerator and the denominator of $\gamma_i$ in \eqref{eq:sinr} by $(1-\alpha_i)$. However if one opts to decrease $\alpha_i$, the transmit power available at the information transceiver of relay $i$ will be further limited [see \eqref{eq:pi}], thus potentially reducing the corresponding data rate $\tau_i$. Similarly, increasing the BS transmit power $P_i$ or the relay transmit power $p_i$ does not necessarily increase the throughput $\tau_i$ of cell $i$. The reason is that $P_i$ and $p_i$ appear in the positive terms in both the numerator and the denominator of $\gamma_i$. This suggests the importance of the resource allocation problem in this context, which will be addressed in the next section.

\section{Joint Resource Optimization Problems for AF Relaying} \label{sec:PF}

In this paper, we aim to devise an optimal tradeoff of all three parameters, transmit power at BSs, $\mathbf{P}$, transmit power at relays, $\mathbf{p}$, and power splitting factor at relays, $\boldsymbol\alpha$, to maximize the performance of the multicell network under consideration. Specifically, we will study the following problems which jointly optimize $(\mathbf{P},\mathbf{p},\boldsymbol\alpha)$ for three different design objectives.

\subsection{Problem (P1): Sum-Rate Maximization} \label{sec:P1}

%The performance of the multicellular network can be optimized by maximizing the total throughput of the network. Since the channel parameters vary from one block to another block, the individual cell can get a fair share of radio resources over a long period of time. Thus, on average, sum throughput maximization problem will maximize the individual cell throughput too.
We assume that $P_\text{max}$ is the maximum power available for transmission at each BS. \redcom{Also, $P_\text{min}$ is the minimum transmit power required at each BS to ensure the activation of energy harvesting circuitry at the relay}. The problem of sum throughput maximization is formulated as follows.
\begin{subequations} \label{eq:P1}
\begin{align}
    \underset{\mathbf{P},\mathbf{p},\boldsymbol\alpha}{\max} \quad
        & \sum_{i=1}^N \tau_i   \label{eq:O1}\\
    \textrm{s.t.} \quad
        &  0 \le \alpha_i \le 1 \;, \;\; \forall i \in \mathcal{N}  \label{eq:C1}\\
        &  \redcom{P_\text{min}} \le P_i \le P_\text{max} \;, \;\; \forall i \in \mathcal{N} \label{eq:C2} \\
        &  0 \le p_i \le \eta \alpha_i \sum_{j=1}^{N} P_j \bh_{j,i}, \;\; \forall i \in \mathcal{N}.\label{eq:C3}
\end{align}
\end{subequations}
In this formulation, \eqref{eq:O1} is the total network throughput whereas \eqref{eq:C1} are the constraints for the power splitting factors for all relays. Also, \eqref{eq:C2} and \eqref{eq:C3} ensure that the transmit powers at the BSs and relays do not exceed the maximum allowable. % Particularly, the throughput function, the throughput, $\tau_i$, is highly nonconvex in heterogeneous multicell network setting. For instance, even, for fixed relay power, $\mathbf{p}$, and fixed relay power splitting factor, $\boldsymbol\alpha$, $\tau_i$, is highly nonconvex in $\mathbf{P}$ due to intercell interference. Thus, optimizing $\mathbf{p}$ and $\boldsymbol\alpha$, in addition to $\mathbf{P}$, is a nontrivial task in order to maximize the total throughput of the network.

%% Explain why \eqref{eq:C3} is non-convex

\subsection{Problem (P2): Max-Min Throughput Fairness}  \label{sec:P2}

In Problem (P1), the network sum-rate is maximized without any consideration given to the throughput actually achieved by the individual users. It might happen that users with more favorable links conditions are allocated with most of the radio resources, leaving nothing for others to fulfill their bare minimum QoS requirements. The latter includes cell-edge users who are the victims of strong intercell interference. In the following, we formulate a max-min fairness problem where the throughput of the most disadvantaged user is maximized.
\begin{subequations} \label{eq:P2}
\begin{align}
    \max_{\mathbf{P},\mathbf{p},\boldsymbol\alpha} \ \min_{i\in\mathcal{N}} \quad
        &  \tau_i   \label{eq:O2}\\
    \textrm{s.t.} \quad
        &  \eqref{eq:C1}-\eqref{eq:C3}. \nonumber% \label{eq:CP2}
\end{align}
\end{subequations}
From the network design perspective, \eqref{eq:P2} can be regarded as the problem of maximizing a common throughput:
\begin{subequations} \label{eq:P2_2}
\begin{align}
    \underset{\mathbf{P},\mathbf{p},\boldsymbol\alpha,\tau}{\max} \quad
        &  \tau   \label{eq:O2_2}\\
    \textrm{s.t.} \quad
        &  \tau_i \ge \tau \ge 0, \quad \forall i\in\mathcal{N}\label{eq:CP2_1}\\
        &  \eqref{eq:C1}-\eqref{eq:C3},  \nonumber%\label{eq:CP2_2}
\end{align}
\end{subequations}
where $\tau$ is an auxiliary variable that denotes the common throughput.
%Similar to sum-throughput maximization problem in \eqref{eq:P1}, the problem in \eqref{eq:P2_2} is highly non-convex because of the non-convexity of the constraints in \eqref{eq:CP2_1} and \eqref{eq:C3}.

%% new variable added

\subsection{Problem (P3): Sum-Power Minimization} \label{sec:P3}

Different from Problems (P1) and (P2), our objective here is to minimize the total transmit power consumption subject to guaranteeing some minimum data throughput $\tau_\text{min}$ for each user:
\begin{subequations} \label{eq:P3}
\begin{align}
    \underset{\mathbf{P},\mathbf{p},\boldsymbol\alpha}{\min} \quad
        &  \sum_{i=1}^N P_i   \label{eq:O3}\\
    \text{s.t.} \quad
        &  \tau_i \ge \tau_\text{min}, \;\; \forall i \in \mathcal{N} \label{eq:CP3_1}\\
        &  \eqref{eq:C1}-\eqref{eq:C3},\nonumber%  \label{eq:CP3_2}
\end{align}
\end{subequations}
This problem is of particular interest for ``green" communications, where one wishes to reduce the environmental impacts of the large-scale deployment of wireless communication networks. At the same time, the performance of all cell-edge users is protected with constraint \eqref{eq:CP3_1}.
% Similar to sum-throughput maximization problem in \eqref{eq:P1}, the problem in \eqref{eq:P3} is highly non-convex because of the non-convexity of the constraints in \eqref{eq:CP3_1} and \eqref{eq:C3}.

All three problems (P1), (P2) and (P3) are \emph{highly nonconvex} in $(\mathbf{P},\mathbf{p},\boldsymbol\alpha)$ because the throughput $\tau_i$ in \eqref{eq:tau} is highly nonconvex in those variables. Even if we fix $\mathbf{p}$ and $\boldsymbol\alpha$ and try to optimize the BS transmit power $\mathbf{P}$ alone, $\tau_i$ would still be highly nonconvex in the remaining variable $\mathbf{P}$ due to the cross-cell interference terms. Simultaneously optimizing $\mathbf{P},\mathbf{p}$ and $\boldsymbol\alpha$  will be much more challenging due to the \emph{nonlinearity} introduced by the cross-multiplying terms, e.g., $P_k p_j \alpha_i$ in \eqref{eq:sinr} and $\alpha_i P_j$ in \eqref{eq:C3}.

To efficiently solve Problems (P1), (P2) and (P3), we propose to adopt the successive convex approximation (SCA) approach \cite{Marks-78-A, PapanEvans-09, Chiang-07-A, Kha-12-A, Ngo-14-A, Razaviyayn-13-A} to transform the original nonconvex problems into a sequence of relaxed convex subproblems. The key steps of the \redcom{generic} SCA approach are summarized in Algorithm~\ref{alg:SCA} for our formulated optimization problems. However, in applying the SCA approach, there remain two key questions: (i) How to perform the approximation in Step 2 in generic Algorithm \ref{alg:SCA}? (ii) Given that the approximation is known, how to prove that the iterative algorithm is convergent to an optimal solution? We will provide the answers for those questions in the following sections. \redcom{Specifically, we will exploit the structure of the formulated problems to propose two types of approximations, one based on GP
programming and the other DC programming. We will demonstrate that with the given objective
functions and constraints, it is possible to apply both approximations to solve the formulated nonconvex problems} \redcom{under the same SCA framework.}

\begin{algorithm}[t]
\caption{Generic Successive Convex Approximation Algorithm}\label{alg:SCA}
\begin{algorithmic}[1]
%\begin{enumerate}
\State Initialize with a feasible solution $(\mathbf{P}^{[0]},\mathbf{p}^{[0]},\boldsymbol\alpha^{[0]})$.
\State At the $m$-th iteration, form a convex subproblem by approximating the nonconcave objective function and constraints of (P1), (P2) and (P3) with some concave function around the previous point $(\mathbf{P}^{[m-1]},\mathbf{p}^{[m-1]},\boldsymbol\alpha^{[m-1]})$.
\State Solve the resulting convex subproblem to obtain an optimal solution $(\mathbf{P}^{[m]},\mathbf{p}^{[m]},\boldsymbol\alpha^{[m]})$ at the $m$-th iteration.
\State Update the approximation parameters in Step 2 for the next iteration.
\State Go back to Step 2 and repeat until $(\mathbf{P},\mathbf{p},\boldsymbol\alpha)$ converges.
\end{algorithmic}
\end{algorithm}

\section{Solutions for AF Relaying: SCA Method Using GP} \label{sec:GP}

To implement Step 2 in Algorithm \ref{alg:SCA}, in this section we will make use of the single condensation approximation method \cite{Chiang-07-A} to form a relaxed geometric program (GP), instead of directly solving the nonconvex Problems (P1), (P2) and (P3). A GP is expressed in the standard form as \cite[p. 161]{Boyd-04-B}:
\begin{subequations}\label{eq:GP}
\begin{align}
    \min_{\mathbf{y}} \quad & f_0(\mathbf{y})\\
    \textrm{s.t.} \quad & f_i(\mathbf{y}) \leq 1, \quad i=1,\ldots,m \\
            \quad & h_{\ell}(\mathbf{y}) = 1, \quad \ell=1,\ldots,M
\end{align}
\end{subequations}
where $f_i(\mathbf{y}), \ i=0,\ldots,m$ are posynomials and $h_{\ell}(\mathbf{y}), \ \ell=1,\ldots,M$ are monomials\footnote{A monomial $\hat{q}(\mathbf{y})$  is defined as $\hat{q}(\mathbf{y}) \triangleq c y_1^{\hat{a}_1} y_2^{\hat{a}_2} \ldots y_n^{\hat{a}_n}$, where $c>0$, ${\mathbf y}=[y_1, y_2, \ldots, y_n]^T \in{\mathbb R}^n_{++}$, and $\hat{\mathbf a}=[\hat{a}_1, \hat{a}_2, \ldots, \hat{a}_n]^T\in{\mathbb R}^n$. A posynomial is a nonnegative sum of monomials. \cite{Boyd-04-B}}. A GP in standard form is a nonlinear and nonconvex optimization problem because posynomials are not convex functions. However, with a logarithmic change of the variables and multiplicative constants, one can easily turn it into an equivalent nonlinear and convex optimization problem (using the property that the log-sum-exp function is convex) \cite{Boyd-04-B, Chiang-07-A}.

\subsection{GP-based Approximated Solution for Problem (P1)} \label{sec:GP_P1}

First, we express the objective function in \eqref{eq:O1} as:
\begin{subequations}\label{eq:P1_O_M}
\begin{align}
     \underset{\mathbf{P},\mathbf{p},\boldsymbol\alpha}{\max} \  \sum_{i=1}^N \frac{1}{2} \log_2 (1 + \gamma_i) & \equiv \underset{\mathbf{P},\mathbf{p},\boldsymbol\alpha}{\max} \ \log_2 \prod_{i=1}^N (1 + \gamma_i)  \label{eq:P1_O_M1} \\
     & \equiv \underset{\mathbf{P},\mathbf{p},\boldsymbol\alpha}{\min} \ \prod_{i=1}^N \frac{1}{1 + \gamma_i}, \label{eq:P1_O_M2}
\end{align}
\end{subequations}
where \eqref{eq:P1_O_M2} follows from \eqref{eq:P1_O_M1} since $\log_2(\cdot)$ is monotonically increasing function. Upon substituting $\gamma_i$ in \eqref{eq:sinr} to \eqref{eq:P1_O_M2} and replacing $1-\alpha_i$ by an auxiliary variable $t_i$, it is shown that Problem (P1) in \eqref{eq:P1} is equivalent to:
\begin{subequations} \label{eq:P1_1}
\begin{align}
    \underset{\mathbf{P},\mathbf{p},\boldsymbol\alpha,\mathbf{t}}{\min} \quad
        & \prod_{i=1}^N \frac{\displaystyle\sum_{j=1,j \ne i}^N \phi_1^{i,j}  P_j p_i t_i + \sum_{j=1}^N \left( \phi_2^{i,j}  P_j t_i + \phi_3^{i,j}  p_j \right) +  \sum_{j=1,j \ne i}^N \sum_{k=1}^N \phi_4^{i,j,k} P_k p_j t_i + 1}{\displaystyle\sum_{j=1}^N \left( \phi_1^{i,j}  P_j p_i t_i + \phi_2^{i,j}  P_j t_i + \phi_3^{i,j}  p_j \right) +  \sum_{j=1,j \ne i}^N \sum_{k=1}^N \phi_4^{i,j,k} P_k p_j t_i +1}\label{eq:O1_1}\\
    \text{s.t.} \quad
        &  t_i + \alpha_i \le 1 \;, \;\; \forall i \in \mathcal{N}  \label{eq:C4}\\
        &  t_i \ge 0 \;, \;\; \forall i \in \mathcal{N} \label{eq:C5} \\
        &   0 \le \frac{p_i}{\eta \alpha_i \sum_{j=1}^{N} P_j \bh_{j,i}} \le 1, \;\; \forall i \in \mathcal{N}.\label{eq:C6}\\
        &   \eqref{eq:C1}, \eqref{eq:C2},\nonumber
\end{align}
\end{subequations}
where $\mathbf{t} \triangleq [t_1,\cdots,t_N]^T$.

It can be seen that \eqref{eq:P1_1} is not yet in the form of \eqref{eq:GP} because \eqref{eq:O1_1} and \eqref{eq:C6} are not posynomials. For notational convenience, let us define:
\begin{align}
    u_i(\mathbf{x}) &\triangleq \sum_{j=1,j \ne i}^N \phi_1^{i,j}  P_j p_i t_i + \sum_{j=1}^N \left( \phi_2^{i,j}  P_j t_i + \phi_3^{i,j}  p_j \right) +  \sum_{j=1,j \ne i}^N \sum_{k=1}^N \phi_4^{i,j,k} P_k p_j t_i + 1,\label{eq:u_i}\\
    v_i(\mathbf{x}) &\triangleq \sum_{j=1}^N \left(\phi_1^{i,j}  P_j p_i t_i + \phi_2^{i,j}  P_j t_i + \phi_3^{i,j}  p_j \right) +  \sum_{j=1,j \ne i}^N \sum_{k=1}^N \phi_4^{i,j,k} P_k p_j t_i + 1,\label{eq:v_i}
\end{align}
where $\mathbf{x} = [ \mathbf{P}^T,\mathbf{p}^T,\mathbf{t}^T ]^T \in\mathbb{R}_+^{3N}$. The objective function in \eqref{eq:O1_1} can then be expressed as:
\begin{align}\label{eq:prod_ui_vi}
     \prod_{i=1}^N  \frac{u_i(\mathbf{x})}{v_i(\mathbf{x})}.
\end{align}
Since $u_i(\mathbf{x})$ and $v_i(\mathbf{x})$ are both posynomials, ${u_i(\mathbf{x})}/{v_i(\mathbf{x})}$ is not necessarily a posynomial, confirming that \eqref{eq:O1_1} is not a posynomial. %Therefore, it is leaving the objective function in \eqref{eq:O1_1} still non-convex.

To transform Problem (P1) into a GP of the form in \eqref{eq:GP}, we would like the objective function \eqref{eq:prod_ui_vi} to be a posynomial. To this end, we propose to apply the single condensation method \cite{Chiang-07-A} and approximate $v_i(\mathbf{x})$ with a monomial $\tilde{v}_i(\mathbf{x})$ as follows. % In this case, ${u_i(\mathbf{x})}/{\tilde{v}_i(\mathbf{x})}$ is a posynomial because it is now the ratio of a posynomial to a monomial. As a result, $\prod_{i=1}^N \left({u_i(\mathbf{x})}/{\tilde{v}_i(\mathbf{x})}\right)$ is also a posynomial because it is a product of many posynomials.
Given the value of $\mathbf{x}^{[m-1]}$ at the $(m-1)$-th iteration, we apply the arithmetic-geometric mean inequality to lower bound $v_i(\mathbf{x})$ at the $m$-th iteration by a monomial $\tilde{v}_i(\mathbf{x})$ as \cite[Lem. 1]{Chiang-07-A}:
\begin{align}\label{eq:app_v}
   v_i(\mathbf{x}) \ge \tilde{v}_i(\mathbf{x}) = & \prod_{j=1}^N \Bigg\{ \left( \frac{v_i(\mathbf{x}^{[m-1]}) P_j p_i t_i }{ P_j^{[m-1]} p_i^{[m-1]} t_i^{[m-1]} } \right)^{\frac{ \phi_1^{i,j} P_j^{[m-1]} p_i^{[m-1]} t_i^{[m-1]}}{v_i(\mathbf{x}^{[m-1]})} } \nonumber\\
   & \qquad \times \left( \frac{v_i(\mathbf{x}^{[m-1]}) P_j t_i }{ P_j^{[m-1]
   } t_i^{[m-1]   } } \right)^{\frac{ \phi_2^{i,j} P_j^{[m-1]} t_i^{[m-1]}}{v_i(\mathbf{x}^{[m-1]})} }\times \left( \frac{v_i(\mathbf{x}^{[m-1]}) p_j }{ p_j^{[m-1]} } \right)^{\frac{ \phi_3^{i,j} p_j^{[m-1]} }{v_i(\mathbf{x}^{[m-1]})} } \Bigg\} \nonumber\\
   & \times v_i(\mathbf{x}^{[m-1]})^{\frac{1}{v_i(\mathbf{x}^{[m-1]})}} \times \prod_{j=1,j\ne i}^N \prod_{k=1}^N \left( \frac{v_i(\mathbf{x}^{[m-1]}) P_k p_j t_i }{ P_k^{[m-1]} p_j^{[m-1]} t_i^{[m-1]} } \right)^{\frac{ \phi_4^{i,j,k} P_k^{[m-1]} p_j^{[m-1]} t_i^{[m-1]}}{v_i(\mathbf{x}^{[m-1]})} }.
\end{align}
It is straightforward to verify that ${v}_i(\mathbf{x}^{[m-1]})=\tilde{v}_i(\mathbf{x}^{[m-1]})$. In fact, $\tilde{v}_i(\mathbf{x})$ is the best local monomial approximation to $v_i(\mathbf{x})$ near $\mathbf{x}^{[m-1]}$ in the sense of the first-order Taylor approximation. With \eqref{eq:app_v}, the objective function ${u_i(\mathbf{x})}/{v_i(\mathbf{x})}$ in \eqref{eq:O1_1} is approximated by ${u_i(\mathbf{x})}/{\tilde{v}_i(\mathbf{x})}$. The latter is a posynomial because $\tilde{v}_i(\mathbf{x})$ is a monomial and the ratio of a posynomial to a monomial is a posynomial. The upper bound $\prod_{i=1}^N \left({u_i(\mathbf{x})}/{\tilde{v}_i(\mathbf{x})}\right)$ of \eqref{eq:prod_ui_vi} is also a posynomial because the product of posynomials is a posynomial. %At the $m$-th iteration), $\prod_{i=1}^N \frac{u_i(\mathbf{x})}{\tilde{v}_i(\mathbf{x})}$ belongs to a class of geometric programming.

Next, we will approximate constraint \eqref{eq:C3} by a posynomial to fit into the GP framework \eqref{eq:GP}. Again, we lower bound posynomial $\eta  \alpha_i \sum_{j=1}^{N} P_j \bh_{j,i}$ by a monomial as \cite[Lem. 1]{Chiang-07-A}:
\begin{align}\label{eq:app_const}
\eta  \alpha_i \sum_{j=1}^{N} P_j \bh_{j,i} \ge w_i(\alpha_i, \mathbf{P}) \triangleq \eta \alpha_i \prod_{j=1}^{N} \left( \frac{P_j \sum_{k=1}^{N} P_k^{[m-1]} \bh_{k,i} }{P_j^{[m-1]}} \right)^\frac{P_j^{[m-1]} \bh_{j,i}}{\sum_{k=1}^{N} P_k^{[m-1]} \bh_{k,i}}.
\end{align}
It is clear that the ratio $p_i/w_i(\alpha_i, \mathbf{P})$ is now a posynomial. Upon substituting \eqref{eq:app_v} and \eqref{eq:app_const} into \eqref{eq:P1_1}, we can formulate an approximated subproblem at the $m$-th iteration for Problem (P1) as follows:
\begin{subequations} \label{eq:P1_2}
\begin{align}
    \underset{\mathbf{x},\boldsymbol\alpha}{\min} \quad
        & \prod_{i=1}^N \frac{u_i(\mathbf{x})}{\tilde{v}_i(\mathbf{x})} \label{eq:O1_2}\\
    \text{s.t.} \quad
        &  0 \le  \frac{p_i}{ w_i(\alpha_i, \mathbf{P})} \le 1 \;, \;\; \forall i \in \mathcal{N} \label{eq:C3_2} \\
        &  \eqref{eq:C1},\eqref{eq:C2},\eqref{eq:C4},\eqref{eq:C5}.\nonumber
\end{align}
\end{subequations}
Comparing with \eqref{eq:GP}, we see that \eqref{eq:P1_2} belongs to the class of a geometric program, i.e., a convex optimization problem. In \eqref{eq:O1_2}, since $v_i(\mathbf{x}) \ge \tilde{v}_i(\mathbf{x})$ [see \eqref{eq:app_v}], we are actually minimizing the upper bound of the original objective function in \eqref{eq:O1_1}. %The accuracy of such an approximation improves in each iteration upon solving an instance of convex subproblem \eqref{eq:P1_2}.
With \eqref{eq:app_const}, constraint \eqref{eq:C3_2} is stricter than \eqref{eq:C3} as:
\begin{align}
\frac{p_i}{\eta  \alpha_i \sum_{j=1}^{N} P_j \bh_{j,i}} \leq \frac{p_i}{w_i(\alpha_i, \mathbf{P})} \leq 1.
\end{align}
%This means that the optimal solution of the approximated problem \eqref{eq:P1_2} always belongs to the feasible set of the original Problem (P1).

\subsection{GP-based Approximated Solution for Problem (P2)}

By substituting $\tau_i$ in \eqref{eq:tau} and carrying out simple algebraic manipulations, constraint \eqref{eq:CP2_1} of Problem (P2) can be rewritten as:
%
%\begin{subequations} \label{eq:P2_3}
\begin{align}
%    \underset{\mathbf{P},\mathbf{p},\boldsymbol\alpha,\tau}{\max} \;
%        &  \tau   \label{eq:O2_2}\\
%    \text{s.t.} \;\;\;\;\;
         \quad \frac{e^{2 \tau \ln 2}}{1 + \gamma_i} \le 1, \;\; \forall i \in \mathcal{N}; \quad \textrm{and} \ \tau \geq 0, \label{eq:CP2_3}
%        &  \tau \ge 0 \label{eq:CP2_4}\\
%        &  \eqref{eq:C1}-\eqref{eq:C3},
\end{align}
%\end{subequations}
%
where $\ln(\cdot)$ denotes the natural logarithm. By introducing the auxiliary variable $\mathbf{t}$ and with $u_i(\mathbf{x})$ and $v_i(\mathbf{x})$ defined in \eqref{eq:u_i}-\eqref{eq:v_i}, it is shown that Problem (P2) is equivalent to:
\begin{subequations} \label{eq:P2_4}
\begin{align}
    \underset{\mathbf{x},\boldsymbol\alpha,\tau}{\max} \quad
        &  \tau   \label{eq:O2_2}\\
    \text{s.t.} \quad
        & \frac{u_i(\mathbf{x}) e^{2 \tau \ln 2}}{v_i(\mathbf{x})} \le 1, \;\; \forall i \in \mathcal{N}  \label{eq:CP2_32} \\
        & \tau \geq 0,\\
        &  \eqref{eq:C1}-\eqref{eq:C3},\eqref{eq:C4},\eqref{eq:C5}.\nonumber
\end{align}
\end{subequations}
As seen, \eqref{eq:P2_4} is not yet in the form of the standard GP \eqref{eq:GP} because constraints \eqref{eq:CP2_32} and \eqref{eq:C3} are not posynomials. Using the similar approach in Sec. \ref{sec:GP_P1}, we can transform \eqref{eq:CP2_32} and \eqref{eq:C3} into posynomials by the approximations in \eqref{eq:app_v} and \eqref{eq:app_const}. The resulting subproblem at the $m$-th iteration of Problem (P2) can be expressed in the standard GP form as:
\begin{subequations} \label{eq:P2_5}
\begin{align}
    \underset{\mathbf{x},\boldsymbol\alpha,\tau}{\max} \quad
        &  \tau   \label{eq:O2_2f}\\
    \text{s.t.} \quad
        &  \frac{u_i(\mathbf{x}) e^{2 \tau \ln 2}}{\tilde{v}_i(\mathbf{x})} \le 1, \;\; \forall i \in \mathcal{N}  \label{eq:CP2_33}\\
        & \tau \geq 0,\\
        &  \eqref{eq:C1},\eqref{eq:C2},\eqref{eq:C4},\eqref{eq:C5},\eqref{eq:C3_2},\nonumber
\end{align}
\end{subequations}
where \eqref{eq:CP2_33} follows directly from \eqref{eq:CP2_32} by replacing $v_i(\mathbf{x})$ with $\tilde{v}_i(\mathbf{x})$ [see in \eqref{eq:app_v}], and \eqref{eq:C3_2} is used in lieu of \eqref{eq:C3}.

\subsection{GP-based Approximated Solution for Problem (P3)}

By introducing an auxiliary variable $\mathbf{t}$ and applying monomial approximation $\tilde{v}_i(\mathbf{x})$ [in \eqref{eq:app_v}] for $v_i(\mathbf{x})$ [in \eqref{eq:v_i}], we can transform the nonconvex constraint \eqref{eq:CP3_1} in Problem (P3) into a posynomial form as:
\begin{align}
    \frac{u_i(\mathbf{x}) e^{2 \tau_\text{min} \ln 2}}{\tilde{v}_i(\mathbf{x})} \le 1.
\end{align}
Again, we use \eqref{eq:C3_2} instead of \eqref{eq:C3} and arrive at the following GP, which is an approximated problem for Problem (P3) at the $m$-th iteration:
\begin{subequations} \label{eq:P3_2}
\begin{align}
    \underset{\mathbf{x},\boldsymbol\alpha}{\min} \quad
        &  \sum_{i=1}^N P_i   \label{eq:O3_2}\\
    \text{s.t.} \quad
        &  \frac{u_i(\mathbf{x}) e^{2 \tau_\text{min} \ln 2}}{\tilde{v}_i(\mathbf{x})} \le 1, \;\; \forall i \in \mathcal{N}  \label{eq:CP2_33b} \\
        &  \eqref{eq:C1},\eqref{eq:C2},\eqref{eq:C4},\eqref{eq:C5},\eqref{eq:C3_2}.\nonumber
\end{align}
\end{subequations}
\subsection{Proposed GP-based SCA Algorithm for Joint Resource Allocation}
It should be noted that GP problems \eqref{eq:P1_2}, \eqref{eq:P2_5} and \eqref{eq:P3_2} are the convex approximations of the original Problems (P1), (P2) and (P3), respectively. In Algorithm \ref{alg:1}, we propose an SCA algorithm in which a (convex) GP is optimally solved at each iteration. %The solution is improved as we iterate by updating the monomial approximation $\tilde{v}_i(\mathbf{x})[m]$.

\begin{algorithm}[!t]\caption{Proposed GP-based SCA Algorithm}\label{alg:1}
  \begin{algorithmic}[1]
  \State Initialize $m := 1$.
  \State Choose a feasible point $\left(\mathbf{x}^{[0]}\triangleq\left(\mathbf{P}^{[0]},\mathbf{p}^{[0]},\mathbf{t}^{[0]}\right); \boldsymbol\alpha^{[0]}\right)$.
  \State Compute the value of $v_i(\mathbf{x}^{[0]}), \ \forall i \in \mathcal{N}$ according to \eqref{eq:v_i}.
  \Repeat
  \State Using $v_i(\mathbf{x}^{[m-1]})$, form the approximate monomial $\tilde{v}_i(\mathbf{x})$ according to \eqref{eq:app_v}.
  \State Using the interior-point method, solve one GP, i.e., \eqref{eq:P1_2} or \eqref{eq:P2_5} or \eqref{eq:P3_2} to find the $m$-th iteration approximated solution $\left(\mathbf{x}^{[m]}\triangleq\left(\mathbf{P}^{[m]},\mathbf{p}^{[m]},\mathbf{t}^{[m]}\right); \boldsymbol\alpha^{[m]}\right)$ for Problem (P1) or (P2) or (P3), respectively.
  \State Compute the value of $v_i(\mathbf{x}^{[m]}), \ \forall i \in \mathcal{N}$ according to \eqref{eq:v_i}.
 \State Set $m := m+1$.
 \Until{Convergence of $\left(\mathbf{x}, \boldsymbol\alpha\right)$ or no further improvement in the objective value \eqref{eq:O1_2} or \eqref{eq:O2_2f} or \eqref{eq:O3_2}}
  \end{algorithmic}
\end{algorithm}

\begin{proposition}\label{Prop_1}
Algorithm \ref{alg:1} generates a sequence of improved feasible solutions that converge to a point $\left(\mathbf{x}^\star, \boldsymbol\alpha^\star\right)$ satisfying the KKT conditions of the original problems (i.e., Problems (P1), (P2) and (P3)).
\end{proposition}

\begin{IEEEproof}
We will prove that Proposition \ref{Prop_1} holds for the case of GP \eqref{eq:P1_2} and its corresponding Problem (P1). The proofs for GP \eqref{eq:P2_5} (hence Problem (P2)) and GP \eqref{eq:P3_2} (hence Problem (P3)) are similar and will be omitted. From \eqref{eq:app_const}, we have that
%\begin{align}
${p_i}\Big/\left(\eta  \alpha_i \sum_{j=1}^{N} P_j \bh_{j,i}\right) \leq {p_i}/{w_i(\alpha_i, \mathbf{P})}$.
%\end{align}
This means that the optimal solution of the approximated problem \eqref{eq:P1_2} always belongs to the {feasible} set of the original Problem (P1).

Next, since $v_i(\mathbf{x}) \ge \tilde{v}_i(\mathbf{x}), \ \forall \mathbf{x}\in\mathbb{R}_+^{3N}$, it follows that:
%\begin{subequations}
\begin{align}\label{eq:conv_p1}
\prod_{i=1}^N \frac{u_i(\mathbf{x}^{[m]})}{v_i(\mathbf{x}^{[m]})} \le \prod_{i=1}^N \frac{u_i(\mathbf{x}^{[m]})}{\tilde{v}_i(\mathbf{x}^{[m]})} = \underset{\mathbf{x}}{\min} \prod_{i=1}^N \frac{u_i(\mathbf{x})}{\tilde{v}_i(\mathbf{x})} \le  \prod_{i=1}^N \frac{u_i(\mathbf{x}^{[m-1]})}{\tilde{v}_i(\mathbf{x}^{[m-1]})} = \prod_{i=1}^N \frac{u_i(\mathbf{x}^{[m-1]})}{v_i(\mathbf{x}^{[m-1]})},
\end{align}
%\end{subequations}
where the last equality holds because $\tilde{v}_i(\mathbf{x}^{[m-1]})={v}_i(\mathbf{x}^{[m-1]})$. As the actual objective value of Problem (P1) is non-increasing after every iteration, Algorithm \ref{alg:1} will eventually {converge} to a point $\left(\mathbf{x}^\star, \boldsymbol\alpha^\star\right)$.

Finally, it can be verified that
\begin{align}
\nabla \left(\frac{u_i(\mathbf{x})}{v_i(\mathbf{x})}\right) \bigg|_{\mathbf{x} = \mathbf{x}^{[m-1]}} &= \nabla\left(\frac{u_i(\mathbf{x})}{\tilde{v}_i(\mathbf{x})}\right) \Bigg|_{\mathbf{x} = \mathbf{x}^{[m-1]}},\label{eq:kkt1_gp}\\
\nabla \left(\frac{p_i}{\eta \alpha_i \sum_{j=1}^{N} P_j \bh_{j,i}}\right) \bigg|_{{\alpha_i = \alpha_i^{[m-1]}; \mathbf{P} =  \mathbf{P}^{[m-1]}}} &=  \nabla \left(\frac{p_i}{w_i(\alpha_i, \mathbf{P})}\right) \Bigg|_{{\alpha_i = \alpha_i^{[m-1]}; \mathbf{P} =  \mathbf{P}^{[m-1]}}},\label{eq:kkt2_gp}
\end{align}
where $\nabla$ denotes the gradient operator. The results in \eqref{eq:kkt1_gp}-\eqref{eq:kkt2_gp} imply that the KKT conditions of the original Problem (P1) will be satisfied after the series of approximations involving GP \eqref{eq:P1_2} converges to the point $\left(\mathbf{x}^\star, \boldsymbol\alpha^\star\right)$. This completes the proof.
\end{IEEEproof}

\section{Solutions for AF Relaying: SCA Method Using DC Programming}\label{sec:DC}

%In the following subsections, we propose DC based solution to solve problems (P1-P3) and achieve a convex subproblem out of the main problems at $m$th iteration. In DC programming, we first resort to write the objective function, \eqref{eq:O1} for problem P1, and the constraints, \eqref{eq:CP2_1} and \eqref{eq:CP3_1} for problems P2 and P3, respectively, as a difference of two concave functions. Then, we approximate the second concave function by an affine function to achieve a convex subproblem at $m$th iteration.

\subsection{DC-based Approximated Solution for Problem (P1)}
In the GP-based approach proposed in Sec. \ref{sec:GP}, we have eliminated the logarithm function in the objective function to form a posynomial [see \eqref{eq:P1_O_M}] and solve the resulting (convex) GP. In the current approach, we propose to keep the logarithm function and rewrite the throughput expression as:
\begin{align}\label{eq:throughput_vu}
    \log_2\left(1+\gamma_i\right) = &\displaystyle\log_2\left(\sum_{j=1}^N \left(\phi_1^{i,j}  P_j p_i (1-\alpha_i) + \phi_2^{i,j}  P_j (1-\alpha_i) + \phi_3^{i,j}  p_j\right) +  \sum_{j=1,j \ne i}^N \sum_{k=1}^N \phi_4^{i,j,k} P_k p_j (1-\alpha_i) +1\right)\nonumber\\
    &  - \log_2\Bigg(\sum_{j=1,j \ne i}^N \phi_1^{i,j}  P_j p_i (1-\alpha_i) + \sum_{j=1}^N \left( \phi_2^{i,j}  P_j (1-\alpha_i) + \phi_3^{i,j}  p_j \right) + \sum_{j=1,j \ne i}^N \sum_{k=1}^N \phi_4^{i,j,k} P_k p_j \nonumber\\
    & \hspace{12.75cm} \times (1-\alpha_i) +1\Bigg)\nonumber\\
    = & \bar{v}_i(\mathbf{x}) - \bar{u}_i(\mathbf{x}),
\end{align}
where we define $\bar{u}_i(\mathbf{x}) \triangleq \log_2(u_i(\mathbf{x}))$ and $\bar{v}_i(\mathbf{x}) \triangleq \log_2(v_i(\mathbf{x}))$ with $u_i(\mathbf{x})$ and $v_i(\mathbf{x})$ given in \eqref{eq:u_i} and \eqref{eq:v_i}, respectively. We also recall that $\mathbf{x} = [\mathbf{P}^T,\mathbf{p}^T,\mathbf{t}^T ]^T \in\mathbb{R}_+^{3N}$, and $\mathbf{t}=\mathbf{1}-\boldsymbol\alpha \in \mathbb{R}^N_+$.

Using the following logarithmic change of variables:
\begin{align}\label{eq:log_var}
    \bP_i \triangleq \ln P_i; \;\; \bp_i \triangleq \ln p_i; \;\; \bt_i \triangleq \ln t_i; % \quad \ba_i \triangleq \ln \alpha_i; \\
    \;\; \bar{\phi}_1^{i,j} \triangleq \ln \phi_1^{i,j}; \;\; \bar{\phi}_2^{i,j} \triangleq \ln \phi_2^{i,j} ; \;\; \bar{\phi}_3^{i,j} \triangleq \ln \phi_3^{i,j}; \;\; \bar{\phi}_4^{i,j,k} \triangleq \ln \phi_4^{i,j,k},
\end{align}
for all $i,j,k \in \mathcal{N}$, we can further write $\bar{u}_i(\cdot)$ and $\bar{v}_i(\cdot)$ in terms of the sums of exponentials in $\bar{\mathbf{x}}$:
%
%\begin{subequations}
\begin{align}\label{eq:bubv}
\bar{u}_i(\bar{\mathbf{x}}) &= \log_2\left(\sum_{j=1,j \ne i}^N  e^{\bP_j + \bp_i + \bt_i + \bar{\phi}_1^{i,j}} + \sum_{j=1}^N \left( e^{\bP_j + \bt_i + \bar{\phi}_2^{i,j}} + e^{\bp_j + \bar{\phi}_3^{i,j}} \right) +  \sum_{j=1,j \ne i}^N \sum_{k=1}^N e^{\bP_k + \bp_j + \bt_i + \bar{\phi}_4^{i,j,k}} + 1\right)\\
\bar{v}_i(\bar{\mathbf{x}}) &= \log_2\left(\sum_{j=1}^N \left(e^{\bP_j + \bp_i + \bt_i + \bar{\phi}_1^{i,j}} + e^{\bP_j + \bt_i + \bar{\phi}_2^{i,j}} + e^{\bp_j + \bar{\phi}_3^{i,j}} \right) +  \sum_{j=1,j \ne i}^N \sum_{k=1}^N e^{\bP_k + \bp_j + \bt_i + \bar{\phi}_4^{i,j,k}} +1\right),
\end{align}
%\end{subequations}
where $\bar{\mathbf{x}} \triangleq [\bar{\mathbf{P}}^T, \bar{\mathbf{p}}^T, \bar{\mathbf{t}}^T ]^T$, $\bar{\mathbf{P}} \triangleq [\bP_1,\hdots,\bP_N]^T$, $ \bar{\mathbf{p}} \triangleq [\bp_1,\hdots,\bp_N]^T$, and $\bar{\mathbf{t}} \triangleq [\bt_1,\hdots,\bt_N]^T$. Since the log-sum-exp function is convex \cite{Boyd-04-B}, both $\bu_i(\bar{\mathbf{x}})$ and $\bv_i(\bar{\mathbf{x}})$ are convex in $\bar{\mathbf{x}}$. However, their difference $\bar{v}_i(\bar{\mathbf{x}})-\bar{u}_i(\bar{\mathbf{x}})=\log_2\left(1+\gamma_i\right)$ in \eqref{eq:throughput_vu} is not necessarily concave.

Using the first-order Taylor series expansion around a given point $\bar{\mathbf{x}}^{[m-1]}$, we propose to approximate $\bv_i(\bar{\mathbf{x}})$ by an affine function as follows \cite{Kha-12-A}:
\begin{align}\label{eq:aff_app}
{ \bv_i(\bar{\mathbf{x}}) } & \approx { \bv_i \left(\bar{\mathbf{x}}^{[m-1]} \right) }  + \left(\nabla \bv_i \left(\bar{\mathbf{x}}^{[m-1]} \right)\right)^T \left( \bar{\mathbf{x}} -  \bar{\mathbf{x}}^{[m-1]} \right),
\end{align}
where the $\ell$-th element of gradient $\nabla \bv_i \left( \bar{\mathbf{x}} \right)$ is given by:
\begin{align}\label{eq:grad}
\nabla^{({\ell})} \bv_i \left( \bar{\mathbf{x}} \right) & = \frac{1}{v_i \left( \bar{\mathbf{x}} \right) \ln 2 }\nonumber\\
& \times \left\{ \begin{array}{l}%\begin{cases}
\displaystyle  e^{\bP_\ell + \bp_i + \bt_i + \bar{\phi}_1^{i,\ell}} + e^{\bP_\ell + \bt_i + \bar{\phi}_2^{i,\ell}}  +  \sum_{j=1,j \ne i}^N  e^{\bP_\ell + \bp_j + \bt_i + \bar{\phi}_4^{i,j,\ell}},   \quad \textrm{if} \ \ell \in \{1,\hdots,N\}  \\
\displaystyle  e^{\bp_i + \bar{\phi}_3^{i,i}}  + \sum_{j=1}^N e^{\bP_j + \bp_i + \bt_i + \bar{\phi}_1^{i,j}},   \quad   \textrm{if} \ \ell = N+i  \\
\displaystyle  e^{\bp_{\ell-N}+\bar{\phi}_3^{i,\ell-N}} +  \sum_{k=1}^N  e^{\bP_k + \bp_{\ell-N} + \bt_i + \bar{\phi}_4^{i,\ell-N,k}},   \quad  \textrm{if} \  \ell \in \{N+1,\hdots,2N\} \setminus \{N+i\}  \\
\displaystyle\sum_{j=1}^N \left( e^{\bP_j + \bp_i + \bt_i + \bar{\phi}_1^{i,j}} +  e^{\bP_j + \bt_i +\bar{\phi}_2^{i,j}} \right) +  \sum_{j=1,j \ne i}^N \sum_{k=1}^N  e^{\bP_k + \bp_j + \bt_i + \bar{\phi}_4^{i,j,k}},  \quad  \textrm{if} \   \ell = 2N+i  \\
0,   \quad    \text{otherwise}.
%\end{cases}
\end{array}
\right.
%\nonumber
\end{align}
With the affine approximation \eqref{eq:aff_app} and the convex function $\bar{u}_i(\bar{\mathbf{x}})$, it is clear that the throughput can now be approximated by a concave function as:
\begin{align}\label{eq:throughput_DC}
\log_2\left(1+\gamma_i\right) \approx {\bv_i \left(\bar{\mathbf{x}}^{[m-1]} \right) }  + \left(\nabla \bv_i \left(\bar{\mathbf{x}}^{[m-1]} \right)\right)^T \left( \bar{\mathbf{x}} -  \bar{\mathbf{x}}^{[m-1]} \right) - \bar{u}_i(\bar{\mathbf{x}}).
\end{align}

By the variable change
\begin{align}\label{eq:log_var_alpha}
 \ba_i \triangleq \ln \alpha_i, \quad \forall i\in\mathcal{N}
\end{align}
and upon denoting $\bar{\boldsymbol\alpha} \triangleq [\ba_1,\hdots,\ba_N]^T$, the nonconvex constraint \eqref{eq:C3} of Problem (P1) can be rewritten as:
\begin{align} \label{eq:C3_2DC}
%e^{\bp_i} &\ge 0 \label{eq:C3_1DC} \\
e^{\bp_i} &\le \eta e^{\ba_i} \sum_{j=1}^{N} e^{\bP_j} \bh_{j,i}.
\end{align}
Applying the arithmetic-geometric inequality, we have that:
\begin{align}\label{eq:app_const_DC_0}
   \sum_{j=1}^{N} e^{\bP_j} \bh_{j,i} \ge \prod_{j=1}^{N} \left( \frac{e^{\bP_j} \bh_{j,i} }{\lambda_{j,i}^{[m-1]}} \right)^{\lambda_{j,i}^{[m-1]}},
\end{align}
where $\mathbf{P}^{[m-1]}$ is a fixed point and
\begin{align}\label{eq:app_const_DC}
    \lambda_{j,i}^{[m-1]} \triangleq \frac{e^{\bP_j^{[m-1]}} \bh_{j,i} }{\displaystyle\sum_{k=1}^N e^{\bP_j^{[m-1]}} \bh_{k,i}}.
\end{align}
As such, \eqref{eq:C3_2DC} can be replaced by a stricter constraint:
%\begin{subequations}
\begin{align}%\label{eq:C3_2DC_simp}
& e^{\bp_i} \le \tilde{w}_i(\ba_i, \bar{\mathbf{P}}) \triangleq \eta e^{\ba_i} \prod_{j=1}^{N} \left( \frac{e^{\bP_j} \bh_{j,i} }{\lambda_{j,i}^{[m-1]}} \right)^{\lambda_{j,i}^{[m-1]}}, \label{eq:C3_2DC_simp1}% \\
\end{align}
%\end{subequations}
which is equivalent to the following affine constraint:
\begin{align}
    \bp_i - \ba_i - \sum_{j=1}^N \lambda_{j,i}^{[m-1]} \bP_j - c_i \le 0, \label{eq:C3_2DC_simp4}
\end{align}
where $c_i \triangleq  \ln \eta + \sum_{j=1}^N \lambda_{j,i}^{[m-1]} \left( \ln \bh_{j,i} - \ln \lambda_{j,i}^{[m-1]}  \right)$ is a constant.

From \eqref{eq:throughput_DC} and \eqref{eq:C3_2DC_simp4}, we now have the following convex optimization problem which gives an approximated solution to Problem (P1) at the $m$-th iteration:
\begin{subequations} \label{eq:P1_4_DC}
\begin{align}
    \underset{\bar{\mathbf{x}}, \bar{\boldsymbol\alpha}}{\max} \quad
        & \sum_{i=1}^N  {\bv_i \left(\bar{\mathbf{x}}^{[m-1]} \right) }  + \left(\nabla \bv_i \left(\bar{\mathbf{x}}^{[m-1]} \right)\right)^T \left( \bar{\mathbf{x}} -  \bar{\mathbf{x}}^{[m-1]} \right) - \bar{u}_i(\bar{\mathbf{x}})  \label{eq:O1_4_DC}\\
    \text{s.t.} \quad
        &  e^{\bt_i} + e^{\ba_i} \le 1, \quad \forall i \in \mathcal{N}\label{eq:C1DC}\\
        &  e^{\ba_i} \le 1, \quad \forall i \in \mathcal{N}  \label{eq:C2DC}\\
        &  e^{\bt_i} \le 1, \quad \forall i \in \mathcal{N}  \label{eq:C2aDC}\\
        &  \redcom{P_\text{min} \le } \;\; e^{\bP_i} \le P_\text{max}, \quad \forall i \in \mathcal{N} \label{eq:C3DC}\\
        & \bp_i - \ba_i - \sum_{j=1}^N \lambda_{j,i}^{[m-1]} \bP_j - c_i \le 0, \label{eq:C4DC}
\end{align}
\end{subequations}
where $\bar{\mathbf{x}}^{[m-1]}$ is known from the $(m-1)$-th iteration.
%Unlike Problem (P1), \eqref{eq:P1_4_DC} is a convex optimization problem.

\subsection{DC-based Approximated Solution for Problems (P2) and (P3)}
In this case, we apply the same logarithmic change of variables in \eqref{eq:log_var} and \eqref{eq:log_var_alpha}. We also make use of the results in \eqref{eq:throughput_DC} and \eqref{eq:C3_2DC_simp4} to show that Problem (P2) in \eqref{eq:P2_2} is approximated by:
\begin{subequations} \label{eq:P2_3_DC}
\begin{align}
    \underset{\bar{\mathbf{x}},\bar{\boldsymbol\alpha},\tau}{\max} \quad
        & \tau  \label{eq:O2_4_DC}  \\
    \text{s.t.} \quad
        & {\bv_i \left(\bar{\mathbf{x}}^{[m-1]} \right) }  + \left(\nabla \bv_i \left(\bar{\mathbf{x}}^{[m-1]} \right)\right)^T \left( \bar{\mathbf{x}} -  \bar{\mathbf{x}}^{[m-1]} \right) - \bar{u}_i(\bar{\mathbf{x}}) \geq 2 \tau \geq 0, \quad \forall i \in \mathcal{N} \label{eq:CP2_DC2} \\
        &  \eqref{eq:C1DC}-\eqref{eq:C4DC}.\nonumber
\end{align}
\end{subequations}
It is clear that \eqref{eq:P2_3_DC} is a convex optimization problem for any given point $\bar{\mathbf{x}}^{[m-1]}$.

By a similar approach, Problem (P3) in \eqref{eq:P3} can be approximated by following convex problem:
\begin{subequations} \label{eq:P3_2_DC}
\begin{align}
    \underset{\bar{\mathbf{x}},\bar{\boldsymbol\alpha}}{\min} \quad
        &  \sum_{i=1}^N P_i   \label{eq:O3_4_DC} \\
    \text{s.t.} \quad
        & {\bv_i \left(\bar{\mathbf{x}}^{[m-1]} \right) }  + \left(\nabla \bv_i \left(\bar{\mathbf{x}}^{[m-1]} \right)\right)^T \left( \bar{\mathbf{x}} -  \bar{\mathbf{x}}^{[m-1]} \right) - \bar{u}_i(\bar{\mathbf{x}}) \geq 2 \tau_\text{min}, \quad \forall i \in \mathcal{N}\\
        &  \eqref{eq:C1DC}-\eqref{eq:C4DC},\nonumber
\end{align}
\end{subequations}
where $\bar{\mathbf{x}}^{[m-1]}$ is known from the $(m-1)$-th iteration.

\subsection{Proposed DC-based SCA Algorithm for Joint Resource Allocation}
In Algorithm \ref{alg:2}, we propose an SCA algorithm in which a convex problem based on the DC approximation is optimally solved at each iteration.
\begin{algorithm}[t] \caption{Proposed DC-based SCA Algorithm}\label{alg:2}
  \begin{algorithmic}[1]
  \State Initialize $m := 1$.
  \State Choose a feasible point $\left(\mathbf{x}^{[0]}\triangleq\left(\mathbf{P}^{[0]},\mathbf{p}^{[0]},\mathbf{t}^{[0]}\right); \boldsymbol\alpha^{[0]}\right)$ and evaluate $\left(\bar{\mathbf{x}}^{[0]}\triangleq\left(\bar{\mathbf{P}}^{[0]},\bar{\mathbf{p}}^{[0]},\bar{\mathbf{t}}^{[0]}\right); \bar{\boldsymbol\alpha}^{[0]}\right)$ using \eqref{eq:log_var} and \eqref{eq:log_var_alpha}.
  \State Compute $\bv_i(\bar{\mathbf{x}}^{[0]})$, $\nabla \log_2 \bv_i \left( \bar{\mathbf{x}}^{[0]} \right)$ and $\lambda_{j,i}^{[0]}, \ \forall i,j \in \mathcal{N}$ using \eqref{eq:bubv}, \eqref{eq:grad} and \eqref{eq:app_const_DC}, respectively.
  \Repeat
  \State Given $\bv_i(\bar{\mathbf{x}}^{[m-1]}), \nabla \log_2 \bv_i \left( \bar{\mathbf{x}}^{[m-1]} \right)$ and $\lambda_{j,i}^{[m-1]}$, form one convex problem, i.e., \eqref{eq:P1_4_DC} or \eqref{eq:P2_3_DC} or \eqref{eq:P3_2_DC}.
  \State Using the interior-point method to solve \eqref{eq:P1_4_DC} or \eqref{eq:P2_3_DC} or \eqref{eq:P3_2_DC} for an approximated solution $\left(\bar{\mathbf{x}}^{[m]} \triangleq \left(\bar{\mathbf{P}}^{[m]}, \bar{\mathbf{p}}^{[m]},\bar{\mathbf{t}}^{[m]}\right); \bar{\boldsymbol\alpha}^{[m]}\right)$ of Problem (P1) or (P2) or (P3) at the $m$-th iteration, respectively.
  %\State Update $\{\mathbf{P}^{[m]},\mathbf{p}^{[m]},\boldsymbol\alpha^{[m]},\mathbf{t}^{[m]}\}$ using \eqref{eq:log_var}.
  \State Update $\bv_i(\bar{\mathbf{x}}^{[m]}), \nabla \log_2 \bv_i \left( \bar{\mathbf{x}}^{[m]} \right)$ and $\lambda_{j,i}^{[m]}, \ \forall i,j \in \mathcal{N}$ using \eqref{eq:bubv}, \eqref{eq:grad} and \eqref{eq:app_const_DC}, respectively.
  \State Set $m := m+1$.
  \Until{Convergence of $\left(\bar{\mathbf{x}}, \bar{\boldsymbol\alpha}\right)$ or no further improvement in the objective value \eqref{eq:O1_4_DC} or \eqref{eq:O2_4_DC} or \eqref{eq:O3_4_DC}}
  \State Recover the optimal solution $\left({\mathbf{x}}^{\star}; \boldsymbol\alpha^{\star}\right)$ from $\left(\bar{\mathbf{x}}^{\star}; \bar{\boldsymbol\alpha}^{\star}\right)$ via \eqref{eq:log_var} and \eqref{eq:log_var_alpha}.
  \end{algorithmic}
\end{algorithm}

%
%\begin{algorithm}[!t]\caption{Proposed GP-based SCA Algorithm}\label{alg:1}
%  \begin{algorithmic}[1]
%  \State Initialize $m := 1$.
%  \State Choose a feasible point $\left(\mathbf{x}^{[0]}\triangleq\left(\mathbf{P}^{[0]},\mathbf{p}^{[0]},\mathbf{t}^{[0]}\right); \boldsymbol\alpha^{[0]}\right)$.
%  \State Compute the value of $v_i(\mathbf{x}^{[0]}), \forall i \in \mathcal{N}$ according to \eqref{eq:v_i}.
%  \Repeat
%  \State Using $v_i(\mathbf{x}^{[m-1]})$, form the approximate monomial $\tilde{v}_i(\mathbf{x})$ according to \eqref{eq:app_v}.
%  \State Using an interior-point method, solve one GP, i.e., \eqref{eq:P1_2} or \eqref{eq:P2_5} or \eqref{eq:P3_2} to find the $m$-th iteration approximated solution $\left(\mathbf{x}^{[m]}\triangleq\left(\mathbf{P}^{[m]},\mathbf{p}^{[m]},\mathbf{t}^{[m]}\right); \boldsymbol\alpha^{[m]}\right)$ for Problem (P1) or (P2) or (P3).
%  \State Compute the value of $v_i(\mathbf{x}^{[m]}), \forall i \in \mathcal{N}$ according to \eqref{eq:v_i}.
% \State Set $m := m+1$.
% \Until{Convergence of $\left(\mathbf{x}, \boldsymbol\alpha\right)$ or no further improvement in the objective function \eqref{eq:O1_2} or \eqref{eq:O2_2f} or \eqref{eq:O3_2}}
%  \end{algorithmic}
%\end{algorithm}

\begin{proposition}\label{Prop_2}
Algorithm \ref{alg:2} generates a sequence of improved feasible solutions that converge to a point $\left(\mathbf{x}^\star; \boldsymbol\alpha^\star\right)$ satisfying the KKT conditions of the original problems (i.e., Problems (P1), (P2) and (P3)).
\end{proposition}

\begin{IEEEproof}
We will prove that Proposition \ref{Prop_2} holds for the case of \eqref{eq:P1_4_DC} and its corresponding Problem (P1). The proofs for  \eqref{eq:P2_3_DC} (hence Problem (P2)) and \eqref{eq:P3_2_DC} (hence Problem (P3)) are similar and will be omitted. From \eqref{eq:app_const_DC_0}, we have that
%\begin{align}
$e^{\bp_i} \Big/ \left( \eta e^{\ba_i} \sum_{j=1}^{N} e^{\bP_j} \bh_{j,i}  \right) \leq e^{\bp_i}/ \tilde{w}_i(\ba_i, \bar{\mathbf{P}})$.
%{\color{red} Please define $\tilde{\bar{w}}_i(\ba_i, \bar{\mathbf{P}})$ to be the right hand side of \eqref{eq:app_const_DC} }.
%\end{align}
Imposing a stricter constraint means that the optimal solution of the approximated problem \eqref{eq:P1_4_DC} always belongs to the {feasible} set of the original Problem (P1).

Because the gradient of the convex function $\bv_i(\bar{\mathbf{x}})$ is its subgradient \cite{Boyd-04-B}, it follows that:
\begin{align}\label{eq:app_v_ineq}
\bv_i(\bar{\mathbf{x}}) \ge
  \bv_i \left(\bar{\mathbf{x}}^{[m-1]} \right)  + \left(\nabla \bv_i \left(\bar{\mathbf{x}}^{[m-1]} \right)\right)^T \left( \bar{\mathbf{x}} -  \bar{\mathbf{x}}^{[m-1]} \right), \quad \forall \mathbf{x}\in\mathbb{R}_+^{3N}.
\end{align}
We now have the following relations for the approximated objective value \eqref{eq:O1_4_DC} at the $m$-th iteration:
\begin{align}\label{eq:conv_p2}
\sum_{i=1}^N  \bv_i (\bar{\mathbf{x}}^{[m]}) - \bu_i(\bar{\mathbf{x}}^{[m]}) & \ge  \sum_{i=1}^N  \bv_i \left(\bar{\mathbf{x}}^{[m-1]} \right)  + \left(\nabla \bv_i^T\left(\bar{\mathbf{x}}^{[m-1]}\right)\right)^T \left( \bar{\mathbf{x}}^{[m]} -  \bar{\mathbf{x}}^{[m-1]} \right) - \bu_i(\bar{\mathbf{x}}^{[m]})\nonumber\\
& = \max_{\bar{\mathbf{x}}} \sum_{i=1}^N  \bv_i \left(\bar{\mathbf{x}}^{[m-1]} \right)  + \left(\nabla \bv_i^T\left(\bar{\mathbf{x}}^{[m-1]}\right)\right)^T \left( \bar{\mathbf{x}} -  \bar{\mathbf{x}}^{[m-1]} \right) - \bu_i(\bar{\mathbf{x}})\nonumber\\
& \ge \sum_{i=1}^N  \bv_i \left(\bar{\mathbf{x}}^{[m-1]} \right)  + \left(\nabla \bv_i^T\left(\bar{\mathbf{x}}^{[m-1]}\right)\right)^T \left( \bar{\mathbf{x}}^{[m-1]} -  \bar{\mathbf{x}}^{[m-1]} \right) - \bu_i(\bar{\mathbf{x}}^{[m-1]})\nonumber\\
& = \sum_{i=1}^N  \bv_i (\bar{\mathbf{x}}^{[m-1]}) - \bu_i(\bar{\mathbf{x}}^{[m-1]})
\end{align}
It is clear \redcom{that} the actual objective value of Problem (P1) is non-decreasing after every iteration. Therefore, Algorithm \ref{alg:2} will eventually {converge} to a point $\left(\mathbf{x}^\star; \boldsymbol\alpha^\star\right)=\left(e^{\bar{\mathbf{x}}^{\star}}; e^{\bar{\boldsymbol\alpha}^{\star}}\right)$.

Finally, it can be verified that
\begin{align}
\nabla \left(\bv_i(\bar{\mathbf{x}}) - \bu_i(\bar{\mathbf{x}}) \right) \bigg|_{\bar{\mathbf{x}} = \bar{\mathbf{x}}^{[m-1]}} = \nabla \left(\bv_i \left(\bar{\mathbf{x}}^{[m-1]} \right) + \left(\nabla \bv_i \left(\bar{\mathbf{x}}^{[m-1]} \right)\right)^T \left( \bar{\mathbf{x}} -  \bar{\mathbf{x}}^{[m-1]} \right) - \bu_i(\bar{\mathbf{x}})\right)  \bigg|_{\bar{\mathbf{x}} = \bar{\mathbf{x}}^{[m-1]}},\label{eq:kkt1_dc}\\
\nabla \left( \frac{e^{\bp_i}}{\eta e^{\ba_i} \sum_{j=1}^{N} e^{\bP_j} \bh_{j,i} } \right) \Bigg|_{{\ba_i = \ba_i^{[m-1]}}; {\bar{\mathbf{P}} =  \bar{\mathbf{P}}^{[m-1]}}}  = \nabla  \left(  \frac{e^{\bp_i}}{ \tilde{w}_i(\ba_i, \bar{\mathbf{P}}) } \right) \Bigg|_{{\ba_i = \ba_i^{[m-1]}}; {\bar{\mathbf{P}} =  \bar{\mathbf{P}}^{[m-1]}}}.\label{eq:kkt2_dc}
\end{align}
The results in \eqref{eq:kkt1_dc}-\eqref{eq:kkt2_dc} imply that the KKT conditions of the original Problem (P1) will be satisfied after the series of approximations involving convex problem \eqref{eq:P1_4_DC} converges to $\left(\bar{\mathbf{x}}^{\star}; \bar{\boldsymbol\alpha}^{\star}\right)$. This completes the proof.
\end{IEEEproof}
%{\color{red} There is one typo in \eqref{eq:C3_2DC_simp4}, where ``$\bP_i - \ba_i \hdots $" is to be replaced by ``$\bp_i - \ba_i \hdots $". Thanks}

\redcom{
\begin{remark}\label{Remark_1}
As discussed in Secs. \ref{sec:GP} and \ref{sec:DC}, we use the SCA framework to propose two different methods, i.e., GP and DC programming, to solve the three problems (P1), (P2), and (P3). In this remark, we present the computational complexity of the two solutions. We first use the big-$\mathcal{O}$ notation to find the computational complexity of the convex subproblems in an iteration \cite{Gahinet-95-A}. To solve problem (P1), the complexity of solving both convex subproblems \eqref{eq:P1_2} (in Algorithm \ref{alg:1}) and \eqref{eq:P1_4_DC} (in Algorithm \ref{alg:2}) is $\mathcal{O} \left( (4N)^3 5N \right)$ because they both have $4N$ optimizing variables and $5N$ constraints. Multiplying this factor by the number of iterations required for convergence, we can obtain the overall computational complexity of Algorithms \ref{alg:1} and \ref{alg:2}. This implies that the order of complexity for both proposed algorithms is the same. Second, in order to compare the exact computational time for the proposed algorithms, we evaluate the CPU execution time \cite{Moller-08-A}. For a fair comparison, the MATLAB codes of the two algorithms are optimized to run on the same computer equipped with Intel Core i7-2670QM, 2.20 GHz processor and 8 GB of RAM. We have observed that the GP-based algorithm is slightly more efficient than DC programming based algorithm, e.g., in solving Problem (P1), Algorithms \ref{alg:1} and \ref{alg:2} on average require $29$sec and $31.5$sec, respectively.
\end{remark}
}

\redcom{
\section{System Model and Proposed Solution for DF Relaying with Variable Timeslot Durations} \label{sec:DF}

In this section, we extend our work to decode-and-forward (DF) relaying. With DF relaying, we have the flexibility to vary the time duration of BS-to-relay and relay-to-user transmissions. In what follows, we will discuss the signal model, sum-rate maximization problem with GP-based solution and the corresponding complexity analysis for DF relaying.

\subsection{Signal Model}
Let $\epsilon T$ define the fraction of the block time used for relay-to-user transmissions. The remaining block time $(1-\epsilon)T$ is used for BS-to-relay energy harvesting and information transmissions. With the signal at the input of information transceiver at relay $i$ in \eqref{eq:yri}, the SINR at the receiver of relay $i$ is given by
\begin{align}\label{SNR_relay_DF}
   \gamma_i^\text{DF-R} = \frac{( 1 - \alpha_i) \bh_{i,i} P_i}{ ( 1 - \alpha_i) \sum_{j=1,j\ne i}^N \bh_{j,i} P_j + \sigma }
\end{align}
The amount of energy harvested at DF relay $i$ is then:
\begin{align}\label{eq:E}
      E_i =  \eta \alpha_i (1-\epsilon) T \sum_{j=1}^{N} P_j \bh_{j,i},
\end{align}
The maximum power available for transmission at DF relay $i$ is $\frac{E_i}{\epsilon T}$, which means that
\begin{align}\label{eq:pi_DF}
      p_i \le \frac{E_i}{\epsilon T} = \eta \alpha_i \frac{1 - \epsilon}{\epsilon} \sum_{j=1}^{N} P_j \bh_{j,i}.
\end{align}
DF relay $i$ will decode the signal from the BS $i$ and forward it to user $i$. Let $\bar{x}_i$ be the decoded version of the signal $x_i$ sent by the BS $i$. The received signal at user $i$ in DF relaying is
\begin{align}\label{eq:yu_DF}
      y_{U_i} =  \frac{g_{i,i}}{\sqrt{ \left( d_{i,i}^g \right)^\beta}} \sqrt{p_i}  \bar{x}_i  +  \sum_{j=1,j \ne i}^{N} \frac{g_{j,i}}{\sqrt{ \left(d_{j,i}^g \right)^\beta}} \sqrt{p_j}  \bar{x}_j + n_i^a.
\end{align}
The SINR at the receiver of user $i$ is thus
\begin{align}\label{SNR_user_DF}
   \gamma_i^\text{DF-U} = \frac{ \bg_{i,i} p_i}{   \sum_{j=1,j\ne i}^N \bg_{j,i} p_j + \sigma }
\end{align}
The achievable throughput in bps/Hz of cell $i$ is then given by
\begin{align}\label{eq:tau_DF}
   \tau_i^\text{DF}(\mathbf{P},\mathbf{p},\boldsymbol\alpha,\epsilon) = \epsilon \log_2 (1 + \gamma_i^\text{DF}),
\end{align}
where $\gamma_i^\text{DF} \triangleq \min \{\gamma_i^\text{DF-R},\gamma_i^\text{DF-U}  \} $.

\subsection{Sum-Rate Maximization Problem and GP-based Solution}
The problem of sum throughput maximization for DF relaying is formulated as follows.
\begin{subequations} \label{eq:P1_DF}
\begin{align}
    \underset{\mathbf{P},\mathbf{p},\boldsymbol\alpha,\epsilon}{\max} \quad
        & \epsilon \sum_{i=1}^N  \log_2 (1 + \min \{\gamma_i^\text{DF-R},\gamma_i^\text{DF-U}  \} )   \label{eq:O1_DF}\\
    \textrm{s.t.} \quad
        &  0 \le \alpha_i \le 1 \;, \;\; \forall i \in \mathcal{N}  \label{eq:C1_DF}\\
        &  P_\text{min} \le P_i \le P_\text{max} \;, \;\; \forall i \in \mathcal{N} \;, \;\; \label{eq:C2_DF}\\
        &  0 \le p_i \le \eta \alpha_i \frac{1 - \epsilon}{\epsilon} \sum_{j=1}^{N} P_j \bh_{j,i}, \;\; \forall i \in \mathcal{N}.\label{eq:C3_DF} \\
        & 0 \le \epsilon \le 1 \;. \;\; \label{eq:Ceps_DF}
\end{align}
\end{subequations}
We will now demonstrate that GP-based SCA approach can be used to solve the nonconvex problem \eqref{eq:P1_DF}\footnote{\redcom{Note that the other problems, i.e., max-min throughput and sum-power minimization, can be similarly formulated and solved for DF relaying. For brevity, they are not presented here.}}.
To transform problem \eqref{eq:P1_DF} into a GP of the form in \eqref{eq:GP}, we first fix $\epsilon$ to find the optimal solution of other parameters and then optimize $\epsilon$ later. By introducing a new auxiliary variable $z_i$, problem \eqref{eq:P1_DF} is equivalently expressed as
\begin{subequations} \label{eq:P1_DF2}
\begin{align}
    \underset{\mathbf{P},\mathbf{p},\boldsymbol\alpha,\mathbf{z}}{\max} \quad
        & \bar\epsilon \sum_{i=1}^N  \log_2 (1 + z_i )   \label{eq:O2_DF}\\
    \textrm{s.t.} \quad
        &  \gamma_i^\text{DF-R} \ge z_i \;, \;\; \forall i \in \mathcal{N} \label{eq:C4_DF} \\
        &  \gamma_i^\text{DF-U} \ge z_i \;, \;\; \forall i \in \mathcal{N} \label{eq:C5_DF} \\
        &  0 \le p_i \le \eta \alpha_i \frac{1 - \bar\epsilon}{\bar\epsilon} \sum_{j=1}^{N} P_j \bh_{j,i}, \;\; \forall i \in \mathcal{N}.\label{eq:C3_DF2} \\
        &  \eqref{eq:C1_DF},\eqref{eq:C2_DF}, \nonumber%
\end{align}
\end{subequations}
where $\mathbf{z} \triangleq [z_1,\hdots,z_N]^T$. The objective function in \eqref{eq:O2_DF} is rewritten as
\begin{align} \label{eq:O2_DF_manip}
       \underset{\mathbf{P},\mathbf{p},\boldsymbol\alpha,\mathbf{z}}{\max} \quad
        & \bar\epsilon \sum_{i=1}^N  \log_2 (1 + z_i ) \equiv \underset{\mathbf{P},\mathbf{p},\boldsymbol\alpha,\mathbf{z}}{\min} \quad \prod_{i=1}^N  \frac{1}{1+z_i}
\end{align}
Next, we approximate the expression $\frac{1}{1+z_i}$ in \eqref{eq:O2_DF_manip} by a posynomial to fit into the GP framework \eqref{eq:GP}. To this end, we lower bound $1+z_i$ by a monomial as \cite[Lem. 1]{Chiang-07-A}:
\begin{align} \label{eq:O2_DF_manip_mon_app}
1+z_i \ge (1+z_i^{[m-1]})^\frac{1}{1+z_i^{[m-1]}} \left( \frac{(1+z_i^{[m-1]}) z_i}{z_i^{[m-1]}} \right)^\frac{z_i^{[m-1]}}{1+z_i^{[m-1]}}.
\end{align}
By using \eqref{eq:O2_DF_manip} and \eqref{eq:O2_DF_manip_mon_app} and ignoring the constant terms, we further reduce \eqref{eq:O2_DF_manip} to
\begin{align} \label{eq:O3_DF_manip}
       \equiv \underset{\mathbf{P},\mathbf{p},\boldsymbol\alpha,\mathbf{z}}{\min} \quad \prod_{i=1}^N  z_i^{-\frac{z_i^{[m-1]}}{1+z_i^{[m-1]}}}
\end{align}
Upon substituting $\gamma_i^\text{DF-R}$ and $\gamma_i^\text{DF-U}$ from \eqref{SNR_relay_DF} and \eqref{SNR_user_DF} into \eqref{eq:P1_DF2}, replacing $1-\alpha_i$ by an auxiliary variable $t_i$, applying arithmetic-geometric mean inequality to lower bound $1+z_i$ and $\sum_{j=1}^{N} P_j \bh_{j,i}$ in \eqref{eq:O2_DF_manip} and \eqref{eq:C3_DF} by monomials, we can formulate an approximated subproblem at the $m$-th iteration for problem \eqref{eq:P1_DF} as follows:
\begin{subequations} \label{eq:P1_DF3}
\begin{align}
    \underset{\mathbf{P},\mathbf{p},\boldsymbol\alpha,\mathbf{t},\mathbf{z}}{\min} \quad & \prod_{i=1}^N z_i^{-\frac{z_i^{[m-1]}}{1+z_i^{[m-1]}}} \label{eq:O3_DF}\\
    \textrm{s.t.} \quad
        & \frac{z_i \left( t_i \sum_{j=1,j\ne i}^N \bh_{j,i} P_j + \sigma  \right)}{ t_i \bh_{i,i} P_i }  \le 1 \;, \;\; \forall i \in \mathcal{N} \label{eq:C4_DF1} \\
        &  \frac{z_i \left( \sum_{j=1,j\ne i}^N \bg_{j,i} p_j + \sigma  \right)}{ \bg_{i,i} p_i } \le 1 \;, \;\; \forall i \in \mathcal{N} \label{eq:C4_DF2} \\
        & 0 \le   \frac{ \bar\epsilon p_i}{ (1-\bar\epsilon) w_i(\alpha_i, \mathbf{P})} \le 1 \;, \;\; \forall i \in \mathcal{N} \label{eq:C5_DF} \\
        &  0 \le t_i \le 1 \;, \;\; \forall i \in \mathcal{N}  \label{eq:C6_DF}\\
        & \alpha_i + t_i \le 1 \;, \;\; \forall i \in \mathcal{N}  \label{eq:C7_DF}\\
        &  \eqref{eq:C1_DF},\eqref{eq:C2_DF}, \nonumber%
\end{align}
\end{subequations}
where $w_i(\alpha_i, \mathbf{P}) \triangleq \eta \alpha_i \prod_{j=1}^{N} \left( \frac{P_j \sum_{k=1}^{N} P_k^{[m-1]} \bh_{k,i} }{P_j^{[m-1]}} \right)^\frac{P_j^{[m-1]} \bh_{j,i}}{\sum_{k=1}^{N} P_k^{[m-1]} \bh_{k,i}}$ is defined in \eqref{eq:app_const}. Compared with \eqref{eq:GP}, problem \eqref{eq:P1_DF3} belongs to the class of geometric programs, i.e., a convex optimization problem. The convergence of the iterative algorithms that solves convex subproblem \eqref{eq:P1_DF3} for DF relaying can be proved using similar steps as stated in Proposition \ref{Prop_1}.

Using the optimized values of $\mathbf{P}$, $\mathbf{p}$, and $\boldsymbol\alpha$, we have to optimize the time fraction $\epsilon$ in the original problem \eqref{eq:P1_DF}. Although \eqref{eq:P1_DF} is linear in $\epsilon$, constraint \eqref{eq:C3_DF} is met with equality at convergence. No further improvement of $\epsilon$ can be achieved by solving \eqref{eq:P1_DF} with the optimized values of $\mathbf{P}$, $\mathbf{p}$, and $\boldsymbol\alpha$. Moreover, constraint \eqref{eq:C3_DF} is not monotonic in $\epsilon$. Hence, the only available option is to apply exhaustive search to find the optimal value of $\epsilon$ in \eqref{eq:P1_DF} for given optimized values of $\mathbf{P}$, $\mathbf{p}$, and $\boldsymbol\alpha$.

\begin{remark}\label{Remark_2}
In the numerical results in Sec. \ref{sec:sim}, we will show that DF relaying with an optimized timeslot fraction results in more than twice the throughput that is otherwise achieved by AF relaying with equal timeslot durations. However, this performance improvement is at the expense of a much higher computational complexity due to the required exhaustive search.
\end{remark}

}

%In the numerical results in Section \ref{sec:sim}, we will see that with equal timeslot assumption for BSs-to-relays and relays-to-users transmission, i.e., $\bar\epsilon = 0.5$, optimization for both AF and DF relaying almost yield similar results. Though, DF relaying provides the flexibility of variable timeslot and further optimization of timeslot fraction, $\epsilon$, improves the performance, however at the expense of increase in computational complexity. For example, if the step size of $0.01$ is used to find the optimal value of $\epsilon$ via linear search, CPU execution time for the convergence of problem \eqref{eq:P1_DF3} in DF relaying with further $\epsilon$ optimization is $477$sec, which is about $18$ times computational complex compared to the execution time of Algorithm \ref{alg:1} in AF relaying ($26.81$sec). Computational efficient optimization of timeslot fraction, $\epsilon$, in DF relaying can be the subject of future research.

\section{Numerical Results}\label{sec:sim}

\begin{figure}[t]
    \centering
    \includegraphics[width=0.5 \textwidth]{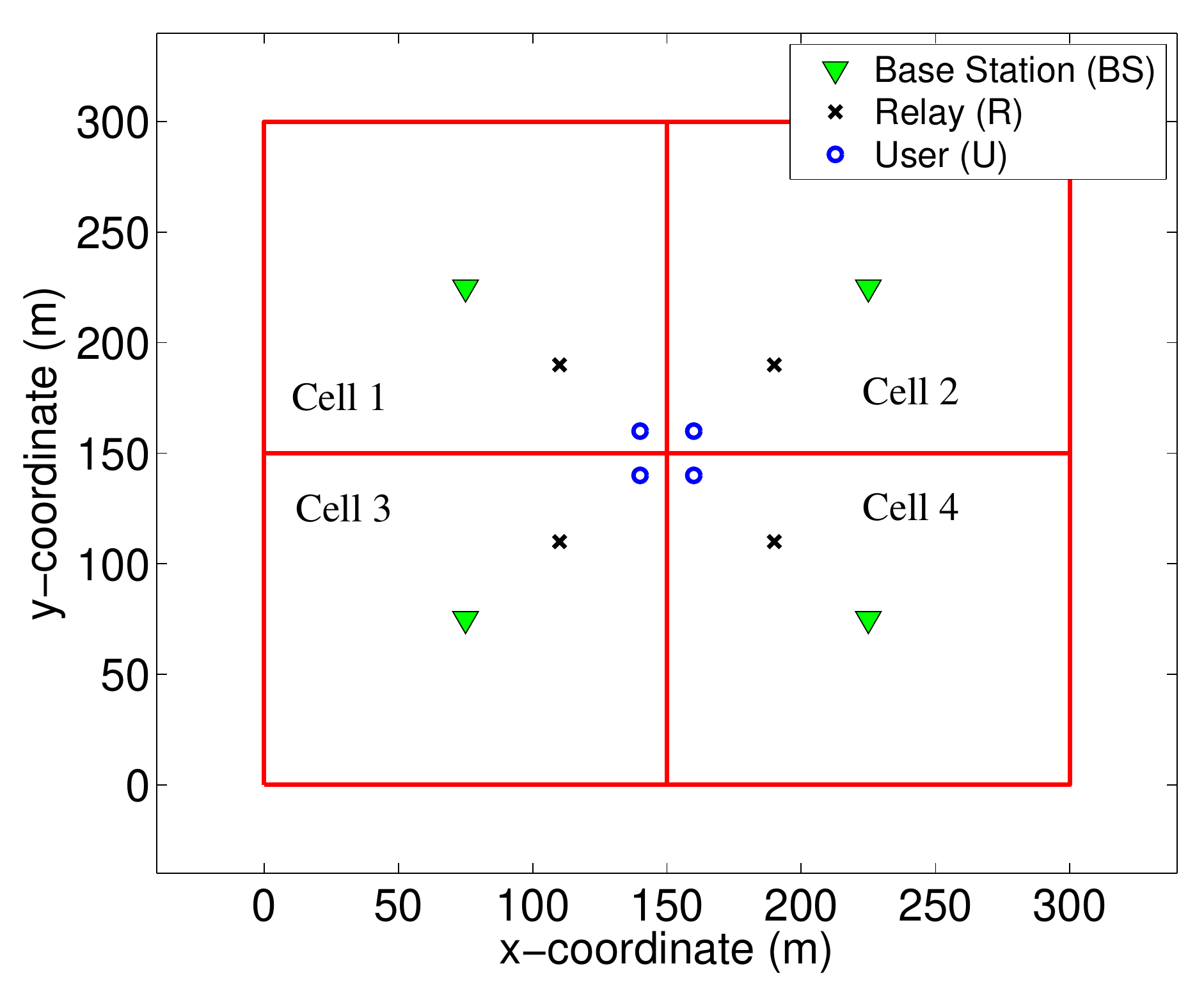}
  \caption{Topology of the multicell network used in the numerical examples.}
    \label{fig:nwtop}
\end{figure}

\redcom{Fig. \ref{fig:nwtop} shows an example multicell network consisting of four $150$m-by-$150$m cells. In each cell, the geographical distance between the servicing BS and its corresponding relay and that between the relay and the cell-edge user is both $35\sqrt{2} \approx 49.5$m, i.e., the relay in each cell is located midway between the BS and the cell-edge user. At the relays, we set the energy harvesting efficiency to $\eta = 0.5$\footnote{\redcom{The value of $\eta$ is typically in the range of $0.4-0.6$ for practical energy harvesting circuits \cite{Lu-14-A}.}}. To model the wireless channels we assume independently and identically distributed block fading. Channel coefficients $h_{i,j}$ and $g_{\bar{j},k}, \ \forall i,j,\bar{j},k$ and $i \ne j$, are circularly symmetric complex Gaussian random variables with zero mean and unit variance. The channel coefficients between the servicing BS and its corresponding relay, i.e., $h_{i,i}$ $\forall$ $i$, are modeled by Rician fading with the Rician factor of $10$ dB. We assume that the randomly-generated values of $h_{i,j}$ and $g_{j,k}$ remain unchanged during each time block where the radio resource allocation process takes place. To model large scale fading, we assume that the path loss exponent is $\beta = 3$. This results in a maximum path loss of $51$ dB between the BS and the associated relay in each cell. In order to activate RF energy harvesting with $\eta = 0.5$ and assuming that the input power at the energy harvesting relay has to be greater than $-25$ dBm \cite{Lu-14-A,Karolak-P-12}\footnote{\redcom{Energy conversion efficiency of around $50\%$ has been reported in the ISM band (900 MHz, 2.4 GHz) with an RF input power of $-25$ dBm and using 13 nm CMOS technology \cite{Lu-14-A,Karolak-P-12}.}}, we set $P_\text{min} = -25+51=26$ dBm.} Using a channel bandwidth of $20$kHz and assuming a noise power density of $-174$dBm/Hz, the total noise power is calculated as $\sigma = -131$dBm \cite{Rep-11}. We initialize the proposed Algorithms \ref{alg:1} and \ref{alg:2} with $P_i^{[0]} = \varsigma P_\text{max}; \ \alpha_i^{[0]} = \varsigma; \ t_i^{[0]} = 1 - \alpha_i^{[0]}; \ p_i^{[0]} = \varsigma \eta \alpha_i^{[0]} \sum_{j=1}^{N} P_j^{[0]} \bh_{j,i}, \ \forall i\in\mathcal{N}$, where $\varsigma$ is a real number taken between $0$ and $1$. To solve each convex problem in Algorithms \ref{alg:1} and \ref{alg:2}, we use CVX, a package for specifying and solving convex programs \cite{cvx, gb08}.

\begin{figure}[t]
\centering
\subfigure[Fixed $\varsigma = 0.5$]{\includegraphics[width=0.48\textwidth]{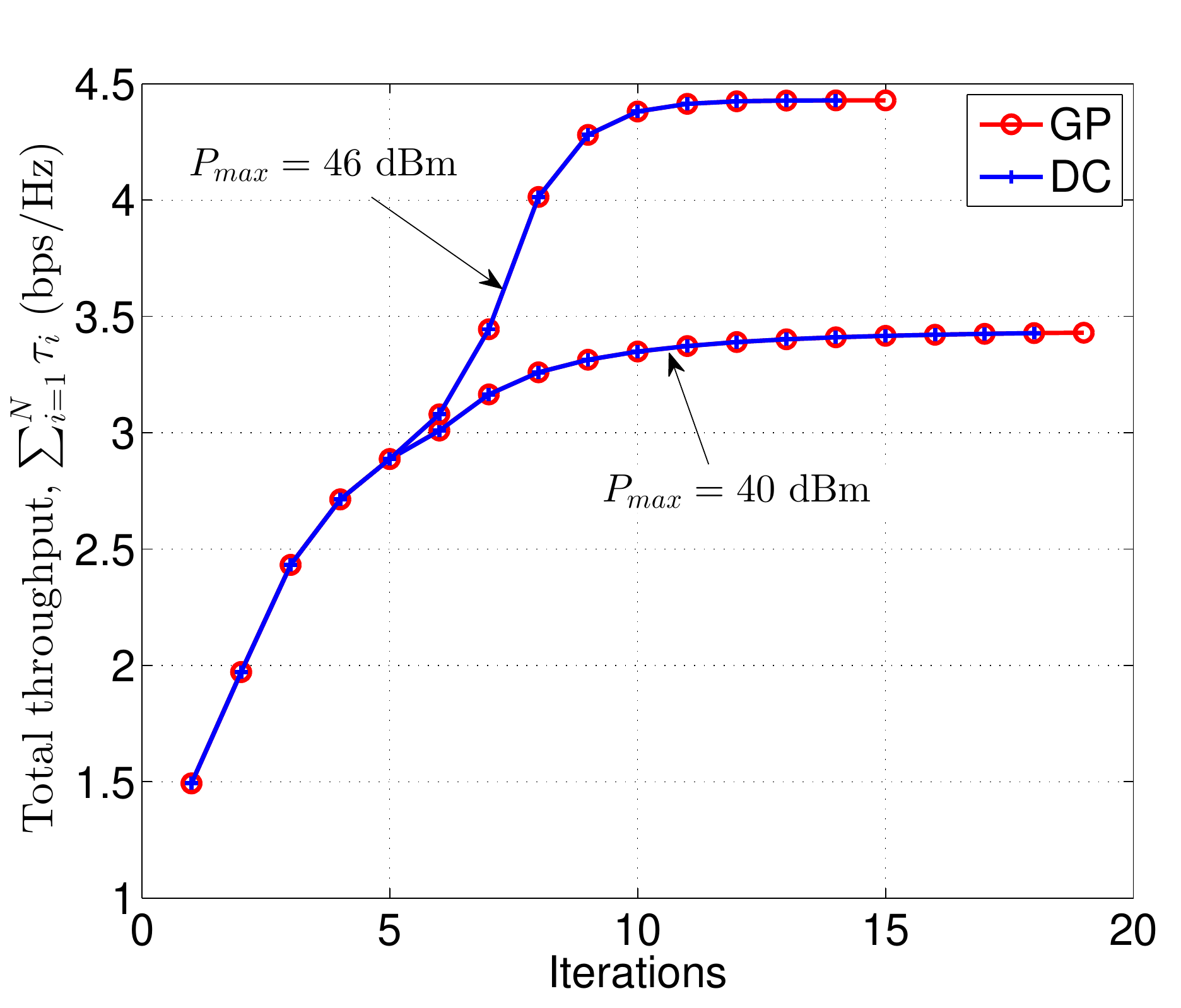}%
\label{fig:P1c}}
%\vfil
\subfigure[Fixed $P_\text{max} = 46$dBm]{\includegraphics[width=0.48\textwidth]{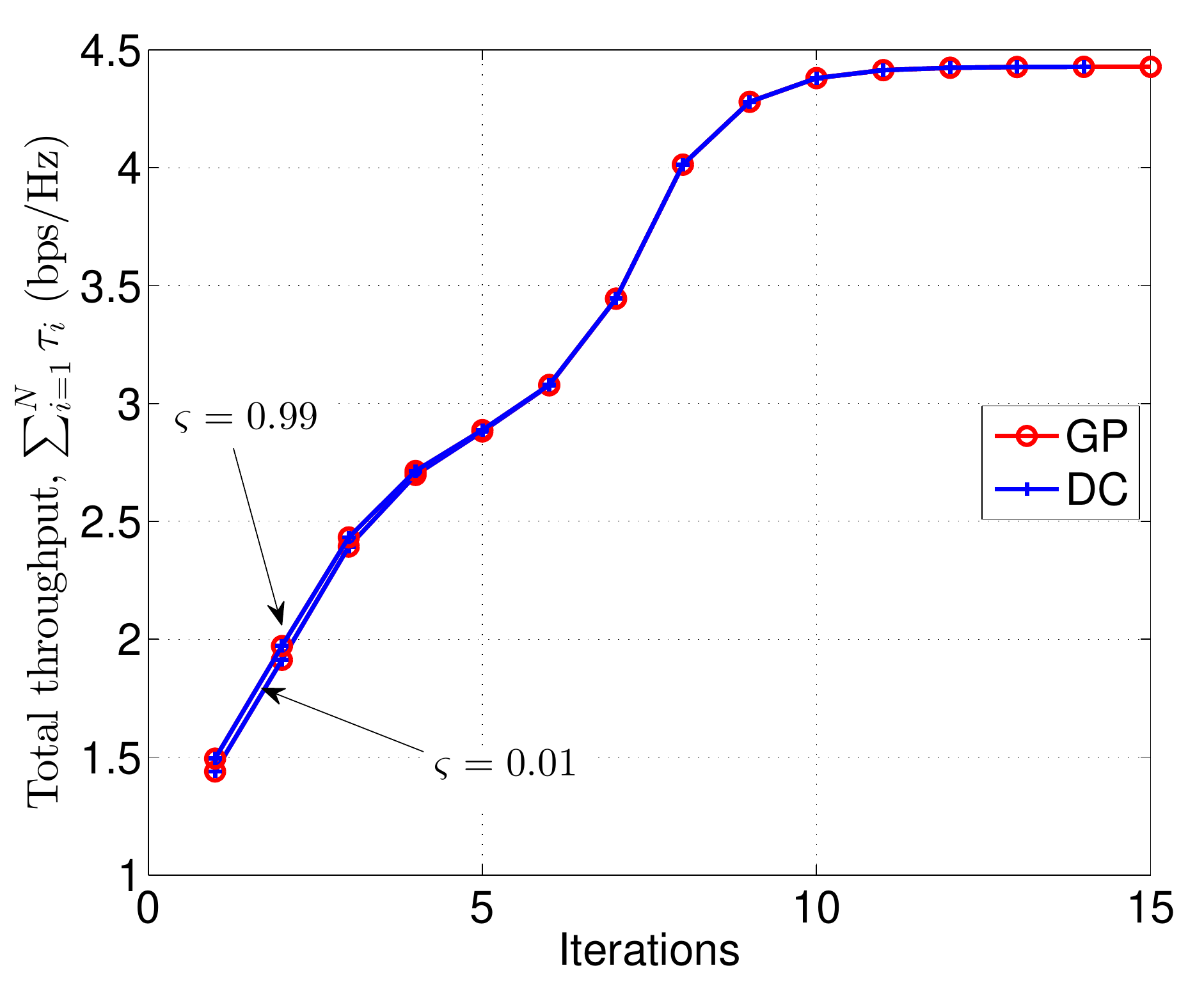}%
\label{fig:P1i}}
\caption{Convergence of Algorithms \ref{alg:1} and \ref{alg:2} in Problem (P1) for AF relaying.}\label{fig:P1ci}
\end{figure}

\subsection{Convergence of the Proposed Algorithms for AF Relaying}

In this subsection, we present numerical results to demonstrate the convergence behavior of the proposed algorithms under different parameter settings.  Regarding Problem (P1), Fig. \ref{fig:P1ci} plots the convergence of the sum throughput $\sum_{i=1}^N \tau_i$ by the proposed solutions. In our simulations, each iteration corresponds to solving of a GP \eqref{eq:P1_2} in Algorithm \ref{alg:1} or a DC program \eqref{eq:P1_4_DC} in Algorithm \ref{alg:2} by CVX. It is clear from Fig. \ref{fig:P1ci} that both algorithms exhibit similar convergence behaviors. In our example, they converge \redcom{within $15$ iterations} and achieve the same optimal throughput. As observed from Fig. \ref{fig:P1c}, the sum rate \redcom{is increased by $28\%$} if we allow a higher BS transmit power budget of $46$dBm instead of \redcom{$40$dBm}. In an interference-limited multicell multiuser network setting, increasing the transmit powers may trigger the `power racing' phenomenon among the users, which in turn adversely affect the total achieved throughput. Our numerical results, on the other hand, confirm that the proposed algorithms effectively manage the strong intercell interference and maximize the network performance. For a fixed power budget $P_{\max}=46$dBm, Fig. \ref{fig:P1i} demonstrates that the final performance of our algorithms is insensitive to the initial points, further suggesting that the solution corresponds to the actual global optimum in our example \cite{Chiang-07-A, PapanEvans-09, Kha-12-A}.

\begin{figure}[t]
\centering
\subfigure[Fixed $\varsigma = 0.5$]{\includegraphics[width=0.48\textwidth]{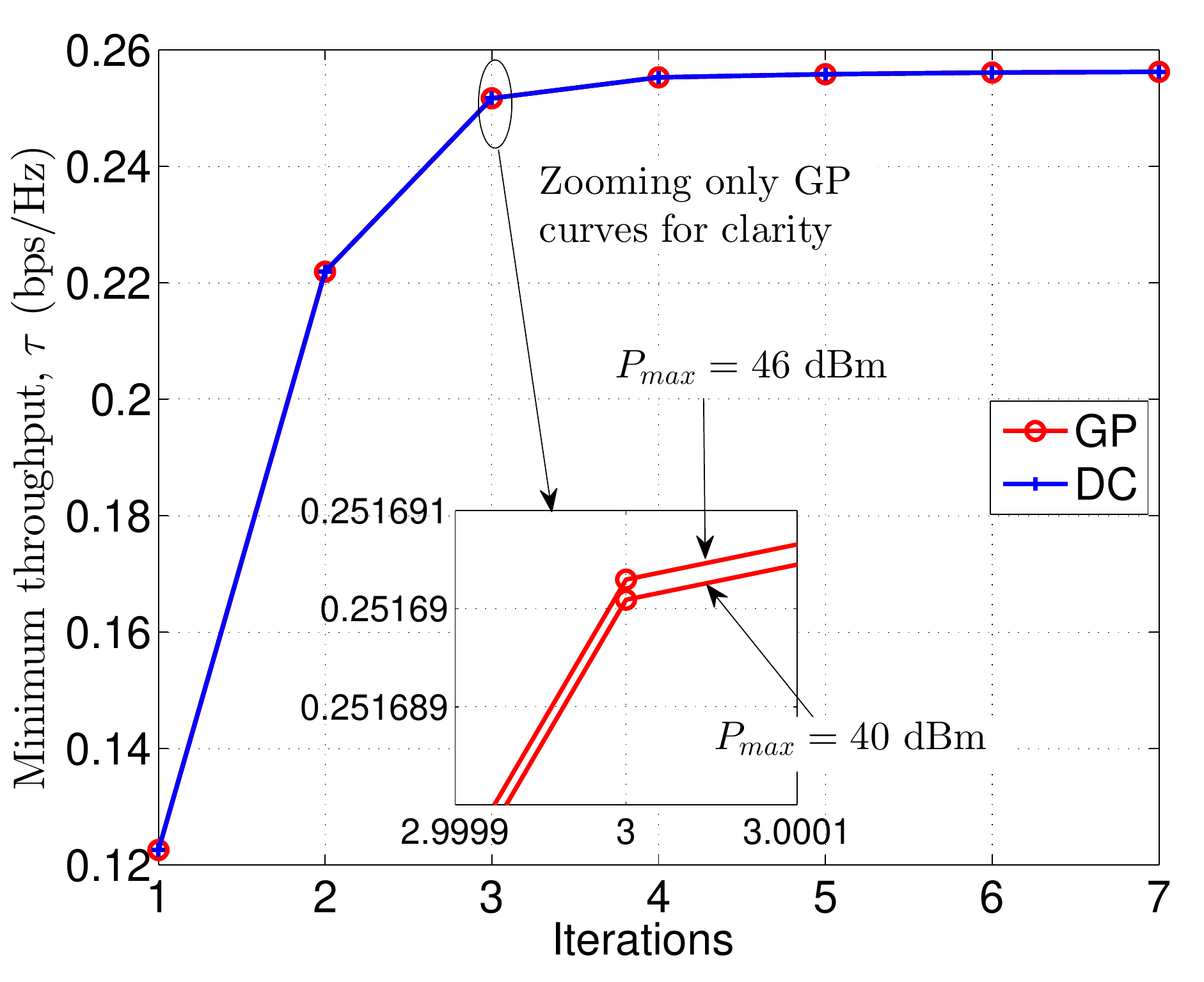}%
\label{fig:P2c}}
%\vfil
\subfigure[Fixed $P_\text{max} = 46$dBm]{\includegraphics[width=0.48\textwidth]{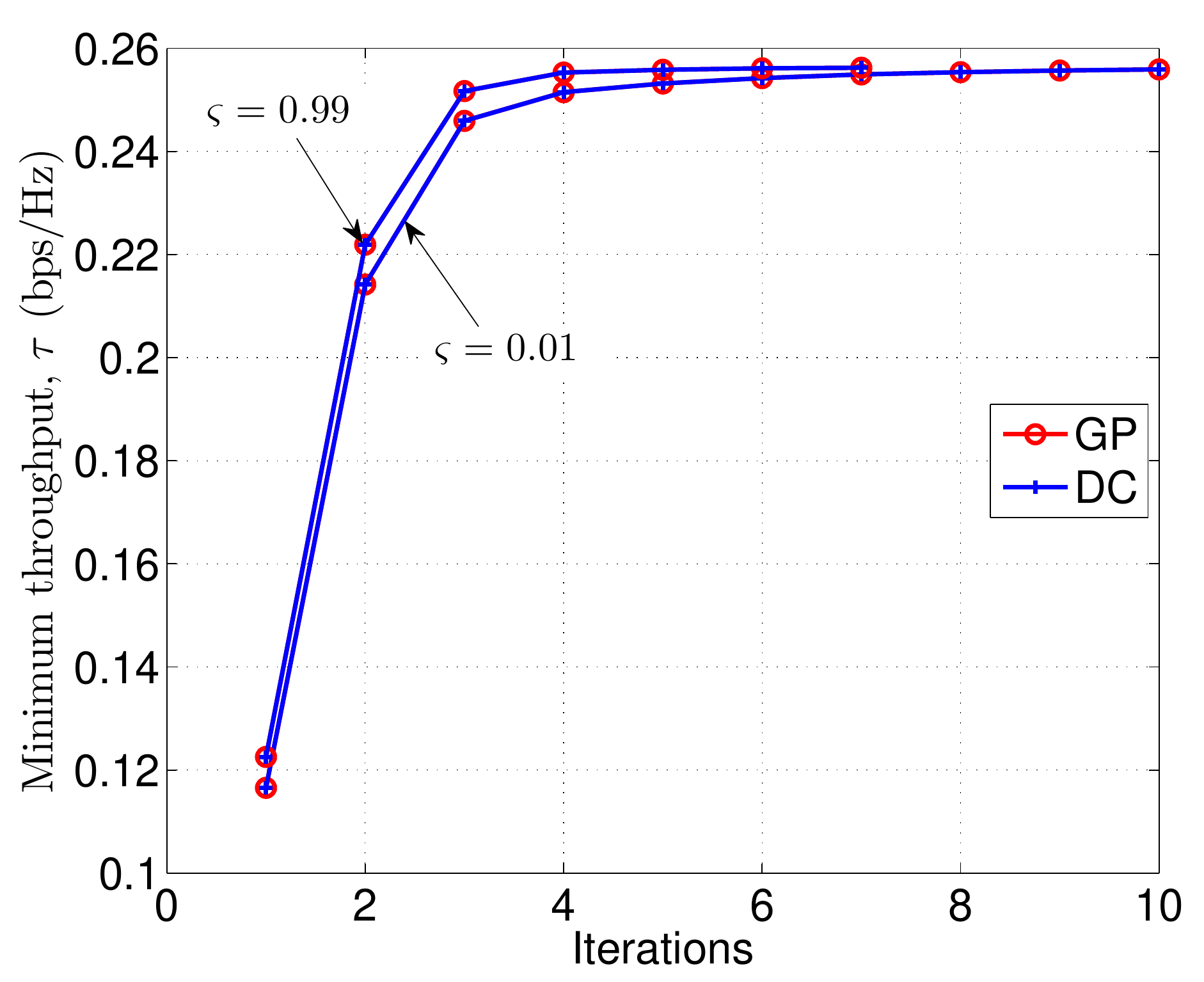}%
\label{fig:P2i}}
\caption{Convergence of Algorithms \ref{alg:1} and \ref{alg:2} in Problem (P2) for AF relaying.\label{fig:P2ci}}
\end{figure}

\begin{figure}[t]
\centering
\subfigure[Fixed $\varsigma = 0.5$]{\includegraphics[width=0.48\textwidth]{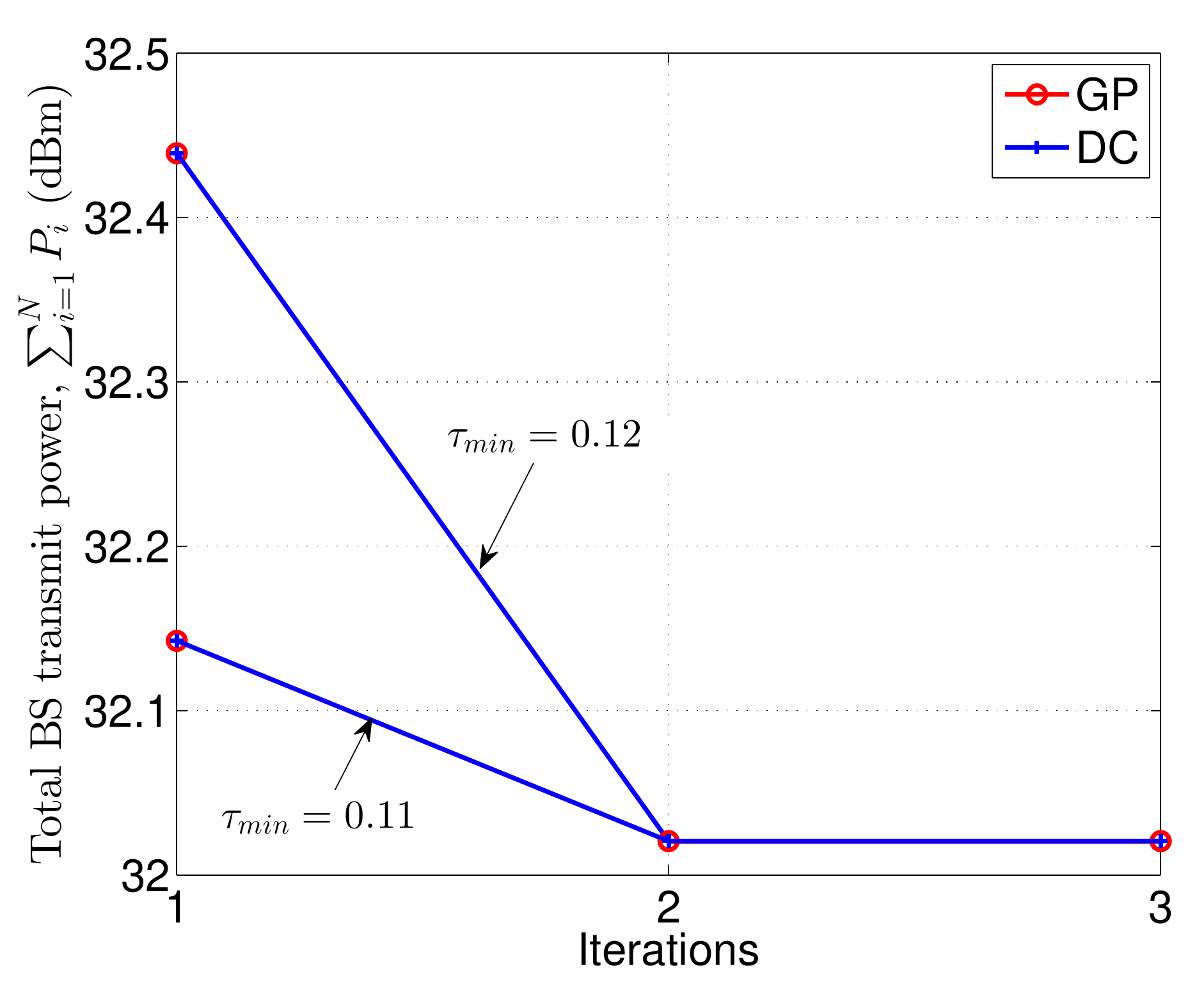}%
\label{fig:P3c}}
%\vfil
\subfigure[Fixed $\tau_\text{min} = 0.12$]{\includegraphics[width=0.48\textwidth]{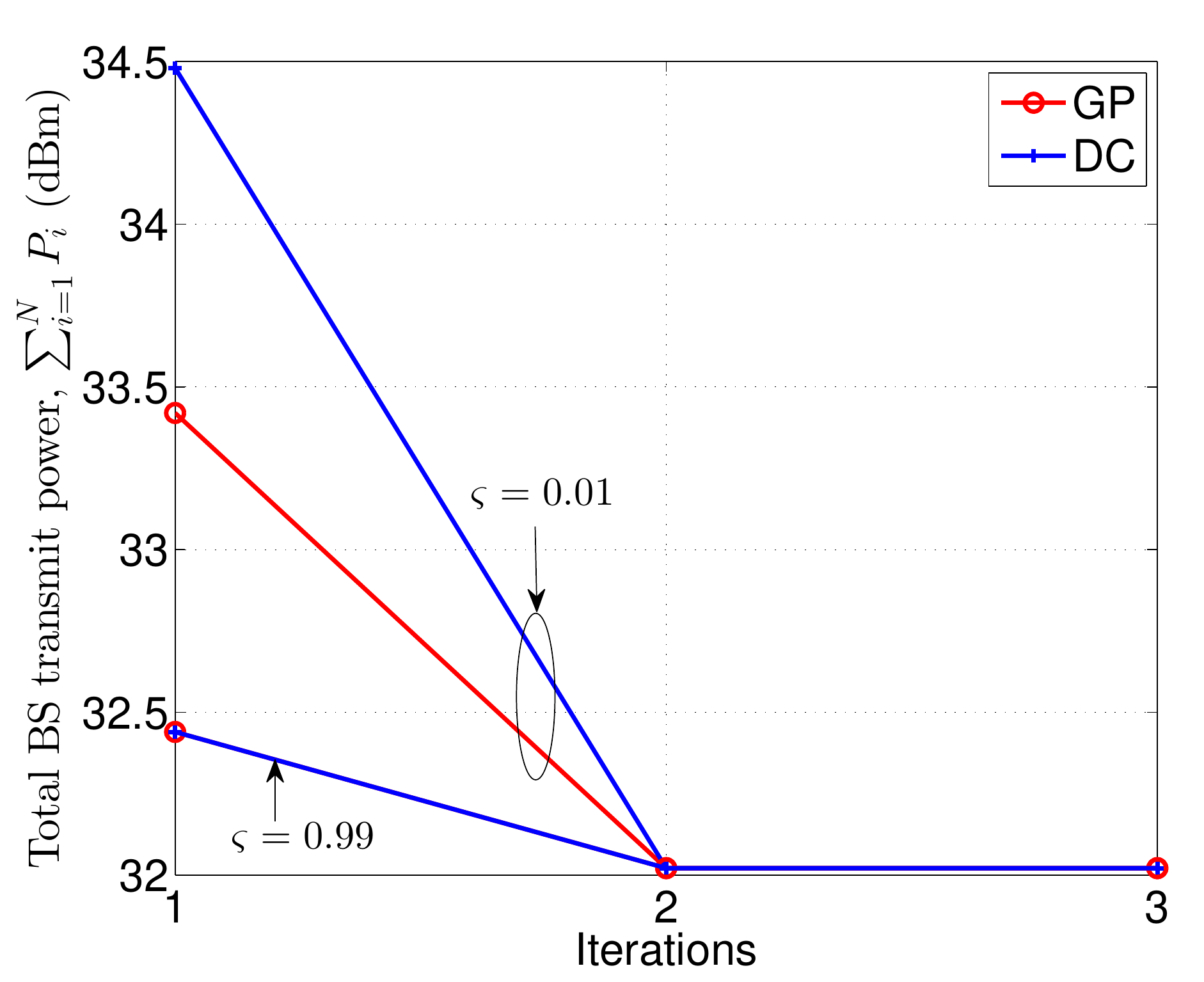}%
\label{fig:P3i}}
\caption{Convergence of Algorithms \ref{alg:1} and \ref{alg:2} in Problem (P3) for AF relaying.\label{fig:P3ci}}
\end{figure}

We demonstrate the performance of our developed algorithms in Figs. \ref{fig:P2ci} and \ref{fig:P3ci} for Problems (P2) and (P3), respectively, \redcom{which plot the convergence of the minimum throughput $\tau$ and total BS transmit power $\sum_{i=1}^N P_i$, respectively}. Again, the proposed algorithms converge quickly to the corresponding optimal values. Different from the results for Problem (P1), increasing $P_\text{max}$ from $40$dBm to $46$dBm in Fig. \ref{fig:P2c} marginally improves the achieved minimum throughput. This signifies the challenge of enhancing the performance of the most disadvantaged user, who is typically located in the cell-edge areas and suffers from the strong intercell interference. In this situation, simply increasing the total allowable transmit power at the BSs would not be helpful. \redcom{On the other hand, Fig. \ref{fig:P3c} verifies that the total required transmit power drops to the minimum value possible, i.e., $N \times P_\text{min} = 32$ dBm for different values of minimum throughput.} Similar to Fig. \ref{fig:P1i}, Figs. \ref{fig:P2i} and \ref{fig:P3i} show that initializing the algorithms with different values of $\varsigma$, again, does not affect the final solutions.

%We can observes from Figs. \ref{fig:P1c}-\ref{fig:P3i} that both algorithms, GP and DC, converge to the same value of the objective functions, sum throughput, common throughput, or total power. Also, it is observed that all optimization parameters, $\mathbf{P}, \mathbf{p}, \boldsymbol\alpha$, converge to the same choices for both algorithms. \textit{It is not possible to justify through exhaustive search or any existing method whether the convergence results are optimal or not. However, according to \cite{Chiang-07-A} and \cite{Kha-12-A}, the successive convex approximation approach, applied in Algorithms 1 and 2, often empirically achieves the globally optimal resource allocation solution.}

\begin{figure}[t]
\centering
\subfigure[Total throughput in Problem (P1)]{\includegraphics[width=0.325\textwidth]{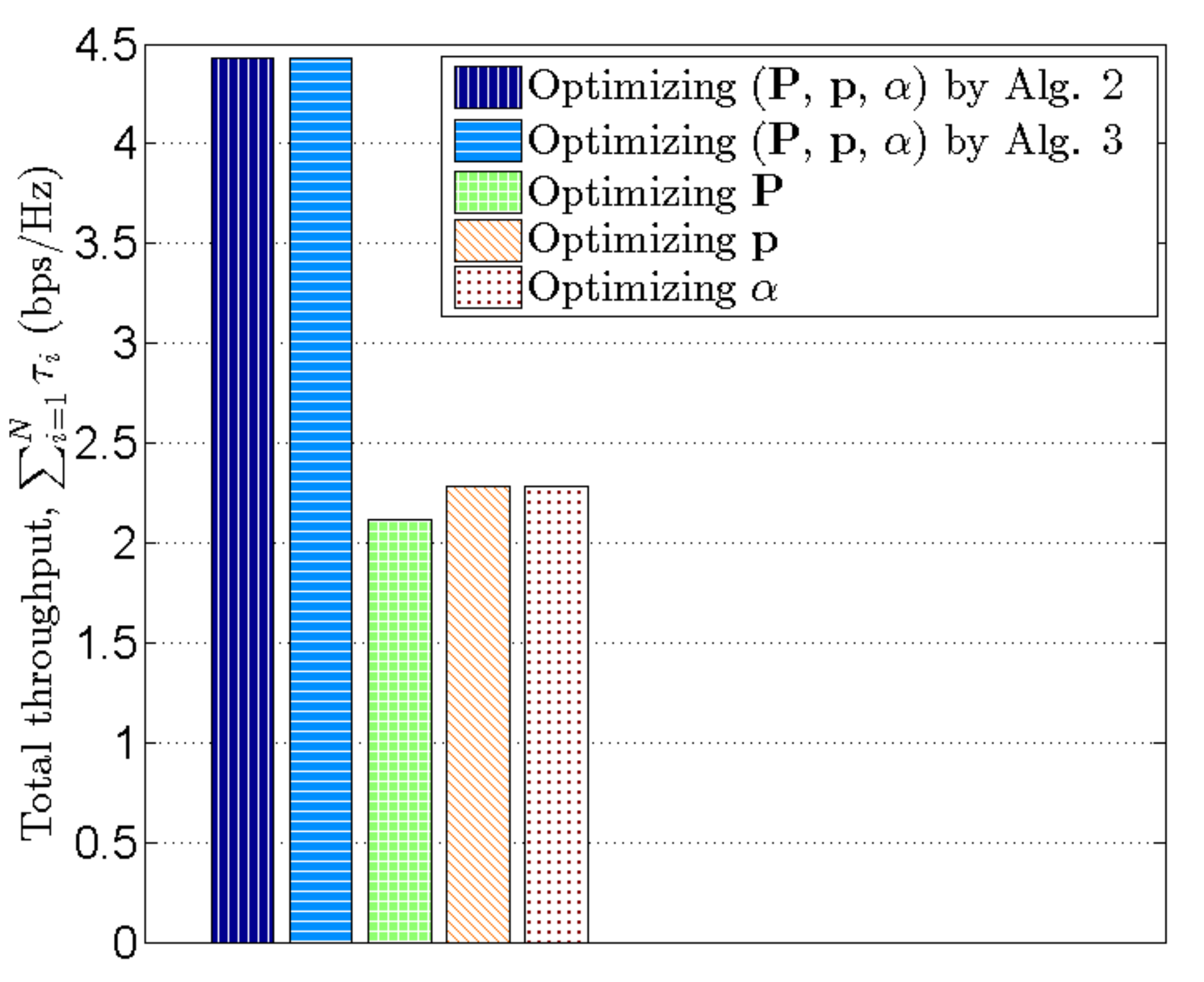}%
\label{fig:P1_jopt}}
%\vfil
\subfigure[Minimum throughput in Problem (P2)]{\includegraphics[width=0.325\textwidth]{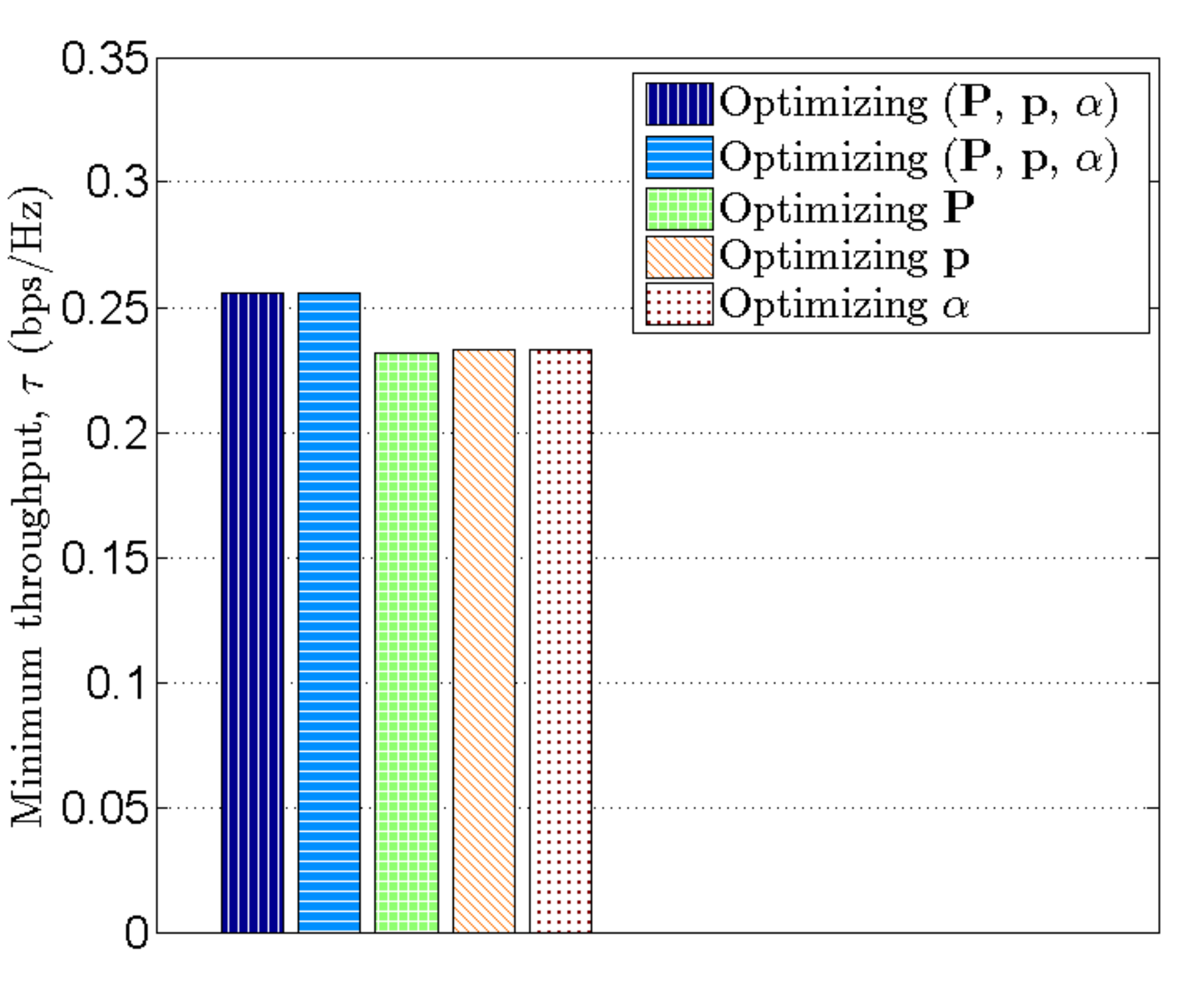}%
\label{fig:P2_jopt}}
\subfigure[Total transmit power in Problem (P3)]{\includegraphics[width=0.325\textwidth]{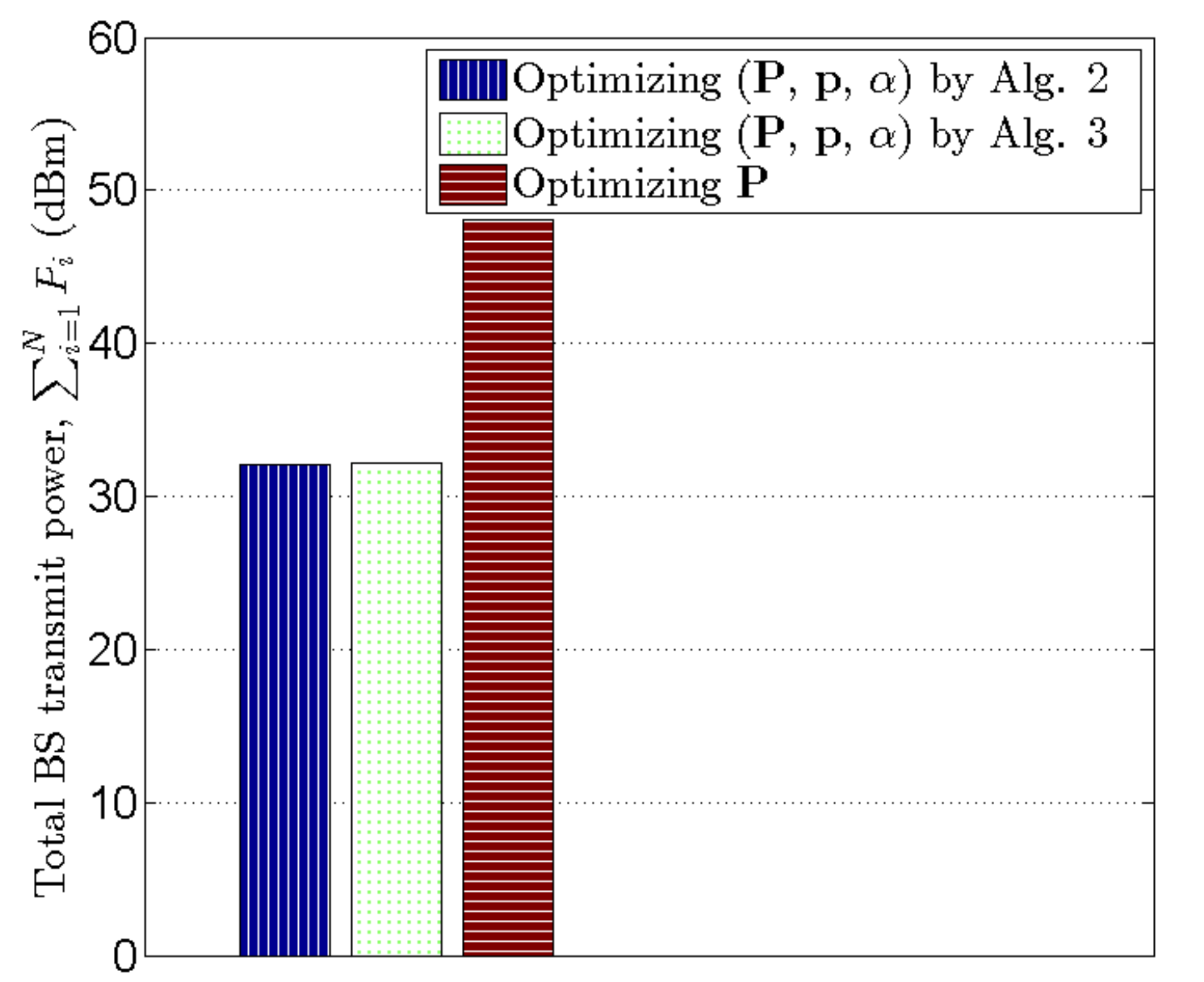}%
\label{fig:P3_jopt}}
\caption{Performance comparison of the proposed joint optimization algorithms and the separate optimization approaches.\label{fig:joint_opt}}
\end{figure}

As seen from Figs.~\ref{fig:P1ci},~\ref{fig:P2ci}~and~\ref{fig:P3ci}, both Algorithms~\ref{alg:1}~and~\ref{alg:2} achieve the same optimal values. However, it is impractical to compare their performance with a globally optimal solution. There is no global optimization approach available in the literature to solve our highly nonconvex optimization problems. A direct exhaustive search would incur a prohibitive computational complexity. It is noteworthy that the works of \cite{Chiang-07-A, PapanEvans-09, Kha-12-A} have shown that the SCA approach often empirically achieves the global optimality in most practical network applications. \redcom{Also since we assume perfect knowledge of CSI at the BSs, the achieved performance corresponds to the theoretical bound that can be obtained. The actual performance with channel estimation errors is out of the scope of this work---a potential future research direction.}

\subsection{Importance of the Proposed Joint Optimization Algorithms for AF Relaying}

Fig.~\ref{fig:joint_opt} demonstrates the advantages of jointly optimizing $(\mathbf{P}, \mathbf{p}, \boldsymbol\alpha)$ as in Algorithms~\ref{alg:1}~and~\ref{alg:2} over optimizing those three parameters individually. In the latter approach, we only optimize one parameter (i.e., $\mathbf{P}$ or $\mathbf{p}$ or $\boldsymbol\alpha$) while fixing the remaining two parameters where applicable as: $P_i=P_\text{max}; \ p_i=\eta \alpha_i \sum_{j=1}^{N} P_j; \ \alpha_i=0.5, \ \forall i \in \mathcal{N}$. Note that for the total power minimization problem (P3), $\mathbf{P}$ is optimized while $\mathbf{p}$ and $\boldsymbol\alpha$ must be fixed. Also in the individual optimization approach, we only present the results of GP-based solutions because both GP and DC approaches achieve similar outcomes.

The results presented in Fig.~\ref{fig:joint_opt} have been averaged over $1,000$ independent simulation runs and we set $\varsigma = 0.5$ and $P_\text{max}=46$dBm. As expected, the proposed joint optimization algorithms outperform the sole optimization approach in all cases. The significant gain is observed in Fig.~\ref{fig:P1_jopt}, where the total throughput is increased by \redcom{$94\%$}. Regarding Problem (P2), Fig.~\ref{fig:P2_jopt} shows that the minimum throughput in Problem (P2) is increased by \redcom{$10\%$} with the proposed Algorithms~\ref{alg:1} and \ref{alg:2}. The performance improvement is less pronounced here. This is because since max-min fairness problem (P2) deals with the most disadvantaged cell-edge user, it is more difficult to support the QoS requirements of such a user compared to only maximizing the overall network performance. Finally, with the minimum throughput $\tau_{\min}=0.12$ required by the most disadvantaged user in Problem (P3), Fig.~\ref{fig:P3_jopt} shows that the proposed algorithms reduce the total BS transmit power by \redcom{$12$ dB i.e., almost $40$ times} over optimizing $\mathbf{P}$ alone.

\begin{figure}[t]
    \centering
    \includegraphics[width=0.5 \textwidth]{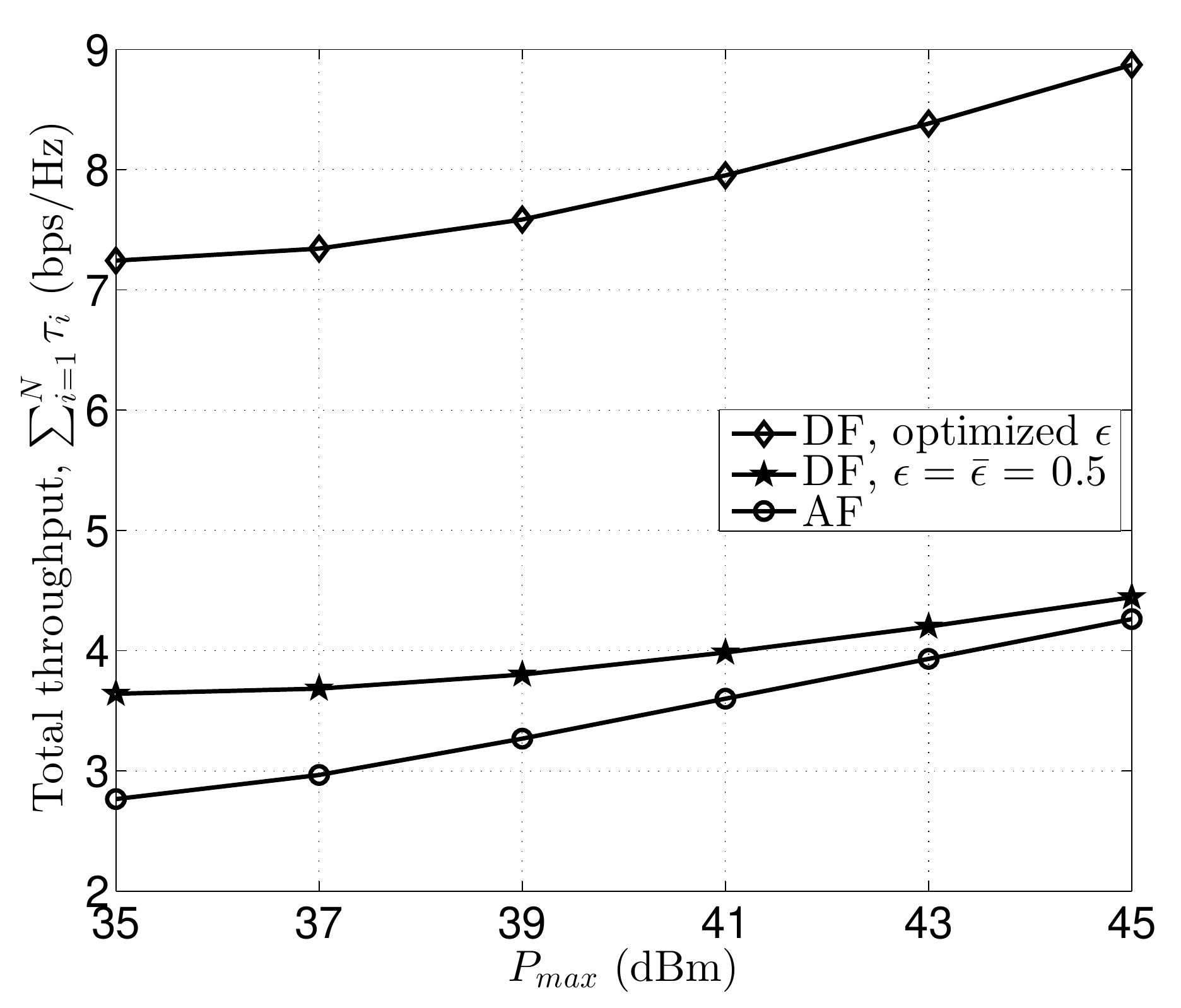}
  \caption{Average sum throughput vs. $P_\text{min} = \{35,37,39,41,43,45\}$ dBm of the proposed joint optimization algorithm (solving Problem (P1)) for AF and DF relaying. The results for DF relaying include both the fixed timeslot case, i.e., $\epsilon = \bar\epsilon = 0.5$ and the optimized $\epsilon$ case.}
    \label{fig:AF_DF}
\end{figure}

\redcom{
\subsection{Comparison of AF and DF Relaying} \label{sec:DF_results}

Fig.~\ref{fig:AF_DF} plots the average sum throughput against different values of $P_\text{max} = \{35,37,39,41,43,45\}$ dBm obtained by the proposed joint optimization algorithm, while solving Problem (P1) for AF and DF relaying. The results for DF relaying include both the equal timeslot case, i.e., $\epsilon = \bar\epsilon = 0.5$ and the optimized $\epsilon$ case. With the equal timeslot assumption for BS-to-relay and relay-to-user transmissions, i.e., $\epsilon = \bar\epsilon = 0.5$, DF relaying increase the throughput by $33\%$ at $P_\text{max} = 35$ dBm. With an optimized value of $\epsilon$, the throughput enhancement can be as high as $170\%$ at $P_{\max} = 35$dBm.

%For the considered range of values of $P_\text{max}$, the optimal value of $\epsilon$ obtained via exhaustive search is around $0.999$. This implies that further improvement in throughput performance is obtained by sparing more time for relay-to-user transmissions.
}

\section{Conclusions and Future Research Directions}\label{sec:conclusions}

In this paper, we have considered the challenging problems for jointly optimizing the BS transmit powers, the relay power splitting factors and the relay transmit powers in a multicell network. It is assumed here that the relay (operating in either AF mode or DF mode) is equipped with a PS receiver architecture that can split the received power in order to scavenge RF energy and to process the information signal from its respective BS. To resolve the highly nonconvex problem formulations, we have proposed SCA algorithms based on geometric programming and DC programming that offer sum-throughput maximization, max-min throughput optimization and sum-power minimization. We have proven that the devised algorithms converge to the solutions that satisfy the KKT conditions of the original nonconvex problems. Illustrative examples have demonstrated the clear advantages of our developed solutions.

\redcom{In case of multiple relays in a cell, two additional problems can be considered for future research (i) in
the first timeslot, beamforming design at the BS toward multiple relays, (ii) in the second time slot,
relay selection to choose which relay to forward the BS message to which users and over which
channel. While these problems are outside the scope of this paper, our proposed solution for the case
of one relay and one user per cell can serve as a first building block toward a joint design in more
general cases.}

\ifCLASSOPTIONpeerreview
\renewcommand{\baselinestretch}{1.0}
\else
\balance
\fi


%\input{Bio}

\end{document}